\documentclass[12pt,preprint]{aastex}
\usepackage{epsfig}

\newcommand{\Hm}{\rm{H}^{-}}
\newcommand{\me}{\rm{e}^-}
\newcommand{\Hp}{\rm{H}^{+}}
\newcommand{\mH}{\rm{H}}
\newcommand{\mHt}{\rm{H}_{2}}
\newcommand{\mHtp}{\rm{H}_{2}^{+}}
\newcommand{\mHtm}{\rm{H}_{2}^{-}}
\newcommand{\hii}{\hbox{H\,{\sc ii}}\,}
\def\simless{\mathbin{\lower 3pt\hbox
   {$\rlap{\raise 5pt\hbox{$\char'074$}}\mathchar''7218$}}}
\def\simgreat{\mathbin{\lower 3pt\hbox
   {$\rlap{\raise 5pt\hbox{$\char'076$}}\mathchar''7218$}}}

\begin{document}

\title{Cosmological Implications of the Uncertainty in {\boldmath H$^-$} 
Destruction Rate Coefficients}

\author{S.~C.~O. Glover}
\affil{Department of Astrophysics, American Museum of Natural History, \\
Central Park West at 79th Street, New York, NY 10024-5192; scog@amnh.org}

\author{D.~W. Savin}
\affil{Columbia Astrophysics Laboratory, Columbia University, \\
550 West 120th Street, New York, NY 10027-6601; savin@astro.columbia.edu}

\and

\author{A.-K. Jappsen}
\affil{Astrophysikalisches Institut Potsdam, An der Sternwarte 16, \\
D-14482 Potsdam, Germany; akjappsen@aip.de}

\begin{abstract}

In primordial gas, molecular hydrogen forms primarily through
associative detachment of $\Hm$ and H, thereby destroying the
$\Hm$. The $\Hm$ anion can also be destroyed by a number of other
reactions, most notably by mutual neutralization with
protons. However, neither the associative detachment nor the mutual
neutralization rate coefficients are well determined: both may be
uncertain by as much as an order of magnitude. This introduces a
corresponding uncertainty into the $\mHt$ formation rate, which may
have cosmological implications.  Here, we examine the effect that
these uncertainties have on the formation of $\mHt$ and the cooling of
protogalactic gas in a variety of situations. We show that the effect
is particularly large for protogalaxies forming in previously ionized
regions, affecting our predictions of whether or not a given
protogalaxy can cool and condense within a Hubble time, and altering
the strength of the ultraviolet background that is required to prevent
collapse.

\end{abstract}


\section{Introduction}

It has long been argued that molecular hydrogen, $\mHt$, must play a
central role in the cooling of primordial gas in the first
protogalaxies.  This is simply because it is the only coolant present
in significant quantities that remains effective at temperatures below
$10^{4} \: {\rm K}$ \citep{sz67,pd68,mst69}. Subsequent numerical work
has only served to confirm this
\citep[e.g.,][]{hutch76,pss83,teg97,mba01,abn02}.  It is now clear
that one of the keys to understanding the earliest episodes of star
formation in the cosmos is a detailed understanding of the chemistry
and thermodynamics of $\mHt$. The chemistry of primordial gas has
recently been reviewed by a number of authors
\citep{aanz97a,gp98,sld98,lsd02}.  These authors find that
considerable chemical complexity is possible despite the limited
number of elements available (essentially only hydrogen, helium, and
lithium, plus isotopic counterparts such as deuterium or $^{3}{\rm
He}$).  But much of this complexity arises from the chemistry of minor
coolants such as ${\rm HD}$ or ${\rm LiH}$, or trace molecules such as
${\rm H}_{3}^{+}$ or ${\rm HeH}^{+}$.  These molecules do not play a
significant role in the chemistry of $\mHt$ \citep{aanz97a,gl01} and
the cooling that they provide is generally unimportant compared to
that coming from $\mHt$.  It is thus usually unnecessary to model their
chemistry (although ${\rm HD}$ cooling can become important in gas
with an atomic hydrogen number density $n_{\mH} > 10^{4} \: {\rm
cm^{-3}}$ and temperature $T < 200 \: {\rm K}$; see \citealt{fpf01} or
\citealt{nu02}).  The omission of these molecules allows substantial
simplifications to be made. In particular, \citet{aanz97a} showed that
the formation and destruction of $\mHt$ can be accurately followed
over a wide range of temperatures and densities with as few as 28
reactions.  In many circumstances, this number can be reduced
further. For instance, four of the reactions deal with the ionization
and recombination of helium, and so play no role in gas in which
helium remains neutral. Furthermore, three of the photochemical
reactions -- the photoionization of $\mH$ and $\mHt$ and the
photodissociation of $\mHt$ by absorption into the continuum of the
Lyman and Werner band systems \citep{ad69} -- require photons with
energies above the Lyman limit and so are important only if a strong
source of hard ultraviolet photons is present.

The key reactions responsible for the formation of $\mHt$ are easily
summarized.  Direct radiative association of atomic hydrogen is
strongly forbidden \citep{gs63}, and so the main gas-phase
pathway by which $\mHt$ is formed makes use of the $\Hm$ ion as an
intermediary.  $\Hm$ is formed by the slow radiative association
reaction
\begin{equation}
 \mH + \me \rightarrow \Hm + \gamma,  \label{hm_ra}
\end{equation}
and is then destroyed by a fast associative detachment reaction with 
atomic hydrogen, forming $\mHt$:
\begin{equation}
 \Hm + \mH \rightarrow \mHt + \me.  \label{hm_ad}
\end{equation}
Molecular hydrogen can also be formed by a similar chain of reactions
involving $\mHtp$: 
\begin{eqnarray}
\mH + \Hp & \rightarrow & \mHtp + \gamma, \\
\mH + \mHtp & \rightarrow & \mHt + \Hp,
\end{eqnarray}
but in most circumstances the $\Hm$ pathway dominates \citep{gl03} as
we now explain.  

Various reactions compete with associative detachment to destroy
$\Hm$, of which the most important are mutual neutralization with
protons
\begin{equation}
\Hm + \Hp \rightarrow \mH + \mH,  \label{hm_mn}
\end{equation}
and photodetachment by infrared and/or optical photons
\begin{equation}
\Hm + \gamma \rightarrow \mH + \me. \label{hm_pd}
\end{equation}
At very high redshifts ($z > 100$), the cosmic microwave background
(CMB) temperature is large enough to produce a substantial
photodetachment rate (Galli \& Palla 1998; Stancil et al.\ 1998)
and so only a few of the $\Hm$
ions survive for long enough to form $\mHt$. At these redshifts,
$\mHt$ formation is dominated by the slower $\mHtp$ pathway. However,
at lower redshifts, photodetachment of $\Hm$ by the CMB rapidly
becomes unimportant; and at the redshifts of interest in this paper
its effect is negligible. The net rate of $\mHt$ formation is
therefore determined by two main factors -- the fractional ionization
of the gas, which controls the rate of the initiating radiative
association (reaction~\ref{hm_ra}), and the fraction of the resulting
$\Hm$ that is destroyed by associative detachment
(reaction~\ref{hm_ad}).  A similar state of affairs exists for
formation of $\mHt$ by the $\mHtp$ pathway, with the main alternative
destruction mechanism in this case being dissociative recombination:
\begin{equation}
 \mHtp + \me \rightarrow \mH + \mH. \label{h2p_dr}
\end{equation}

Once formed, $\mHt$ can be destroyed by charge transfer with $\Hp$, 
\begin{equation}
\mHt + \Hp \rightarrow \mHtp + \mH, \label{h2_ct}
\end{equation}
by collisional dissociation by free electrons or atomic
hydrogen\footnote{Dissociation due to collisions with molecular
hydrogen is also possible, but collisions with atomic hydrogen
dominate due to the fact that the molecular fraction of primordial gas
is small (typically, $n_{\mHt} < 5 \times 10^{-3} n_{\mH}$; see Susa
et al.\ 1998, Nishi \& Susa 1999, Oh \& Haiman 2002 for a detailed
discussion of why this is the case).}
\begin{eqnarray}
\mHt + \me & \rightarrow & \mH + \mH + \me, \\
\mHt + \mH & \rightarrow & \mH + \mH + \mH,
\end{eqnarray}
or by photodissociation via the Solomon process \citep{sw67}
\begin{equation}
\mHt + \gamma \rightarrow \mH + \mH.
\end{equation}

In addition to the reactions listed above, several others are required
to complete our chemical network. Collisional ionization
of $\mH$ by electrons 
\begin{eqnarray}
\mH + \me & \rightarrow & \Hp + \me + \me
\end{eqnarray}
and radiative recombination of $\Hp$
\begin{eqnarray}
\Hp + \me & \rightarrow & \mH + \gamma,
\end{eqnarray}
must be included if we wish to model the evolution of the fractional 
ionization of the gas correctly. The other reactions include the
dissociative recombination of $\mHtp$ (reaction~\ref{h2p_dr} above)
together with 
\begin{eqnarray}
\Hm + \me & \rightarrow & \mH + \me + \me, \\
\Hm + \mH & \rightarrow & \mH + \mH + \me, \\
\Hm + \Hp & \rightarrow & \mHtp + \me, \\
\mHtp + \gamma & \rightarrow & \mH + \Hp.
\end{eqnarray}
These reactions all play a role in regulating either the $\Hm$ or the 
$\mHtp$ abundance, and so affect $\mHt$ formation. However, it should be 
noted that in most circumstances, the influence of these reactions on 
the $\mHt$ formation rate is relatively small.

There is considerable variation in the accuracy with which the rates
and rate coefficients of the reactions in this network are
known. Some, such as the radiative association reaction that forms
$\Hm$ (reaction~\ref{hm_ra}), have rate coefficients which have been
determined to a high level of accuracy \citep[see the discussion
in][]{aanz97a}. Others, however, such as the $\mHt$ charge transfer
reaction (reaction~\ref{h2_ct}) discussed by \citet{skhs04}, have rate
coefficients which are considerably more uncertain. In this paper, we
are concerned with two reactions for which a large degree of
uncertainty exists: the destruction of $\Hm$ by associative detachment
with $\mH$ (reaction~\ref{hm_ad}) and by mutual neutralization with
$\Hp$ (reaction~\ref{hm_mn}). In situations where the photodetachment
of $\Hm$ (reaction~\ref{hm_pd}) is unimportant, it is the competition
between these two reactions that determines what fraction of the $\Hm$
formed by reaction~\ref{hm_ra} goes on to form $\mHt$.  So substantial
variations in the rate coefficients of these reactions can result in
major variations in the $\mHt$ formation rate. This is turn may have
significant consequences for the cooling of primordial gas in
situations of cosmological importance. It is therefore important to
determine the extent to which the existing uncertainties in the
associative detachment and mutual neutralization rate coefficients
affect the conclusions that we can draw regarding the formation of
$\mHt$ in primordial gas.

To investigate this, we have performed several sets of simulations of
protogalactic collapse in which we have varied the two rate
coefficients over a range of plausible values, in an effort to see how
sensitive is the outcome for various initial conditions to the
particular collision data chosen. In Section~\ref{rates} of this
paper, we review the current state of knowledge regarding the values
of the associative detachment and mutual neutralization rate
coefficients.  In Section~\ref{code} we describe the code used to
perform our simulations. The initial conditions used in our
simulations are discussed in Section~\ref{init}, and the results of
the simulations are presented in Section~\ref{result}.  We conclude
with a brief discussion in Section~\ref{discuss}.

\section{Data for Mutual Neutralization and Associative Detachment}
\label{rates}

\begin{figure}
\centering
\epsfig{figure=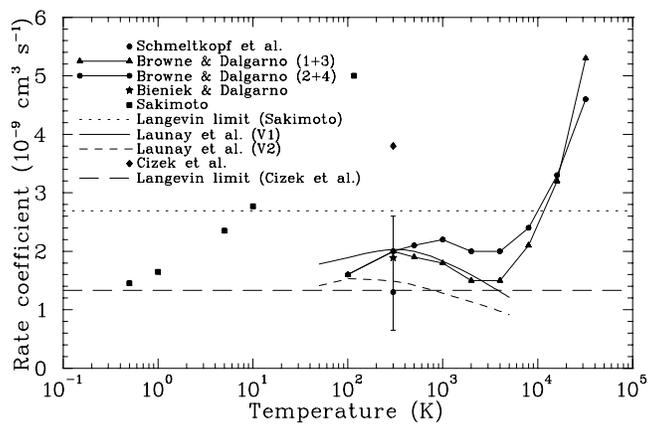,width=20pc,angle=0,clip=}
\caption{Published experimental and theoretical rate coefficients for
the associative detachment reaction $\Hm + \mH \rightarrow \mHtm 
\rightarrow \mHt + \me$ at temperatures relevant for early Universe 
chemistry.}
\label{ad}
\end{figure}

\begin{figure}
\centering
\epsfig{figure=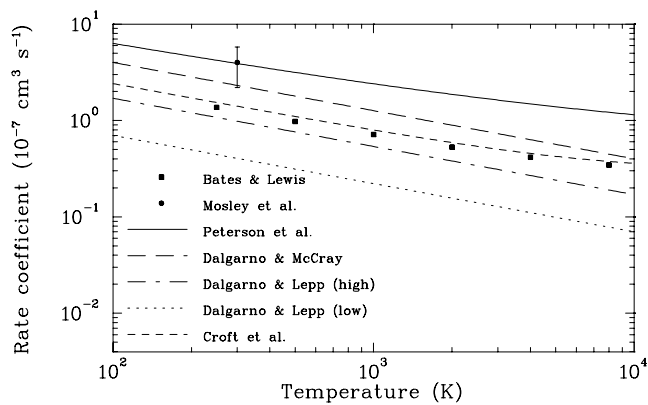,width=20pc,angle=0,clip=}
\caption{Published theoretical and experimental rate coefficients
for the mutual neutralization reaction $\Hm + \Hp \rightarrow \mH + \mH$
at temperatures relevant for early Universe chemistry.}
\label{mn}
\end{figure}

Here we review the published theoretical and experimental work for the
associative detachment reaction~\ref{hm_ad} and the mutual
neutralization reaction~\ref{hm_mn} at temperatures $T$ and collision
energies $E$ relevant for cosmology.  In specific, we focus on work at
$T \lesssim 10^4 \: {\rm K}$ or $E \lesssim 1 \: {\rm eV}$.  The
corresponding published rate coefficients for reactions~\ref{hm_ad}
and \ref{hm_mn} are shown in Figs.~\ref{ad} and \ref{mn},
respectively.

\subsection{Associative Detachment}

The only measurement at cosmological temperatures of the associative
detachment reaction~\ref{hm_ad} has been carried out by Schmeltekopf,
Fehsenfeld, \& Ferguson (1967).  They used a fast-flowing afterglow
system and claimed their measured rate coefficient is accurate to
within a factor of 2.  This was quickly followed with a semi-classical
theoretical calculation by \citet{db67} using a simple extension on
the theory for radiative association.  \citet{bd69} subsequently
revised their earlier calculations using new potentials for the
intermediate $\mHtm$ molecular anion.  They presented results for
their potentials (1+3) and (2+4).  Both of these results from Browne
\& Dalgarno are shown in Fig.~\ref{ad}.  All of the above theory takes
into account the lowest two electronic states of $\mHtm$, namely the
attractive ${\rm X}\ ^2\Sigma_u^+$ state and the repulsive ${\rm B}\
^2\Sigma_g^+$ state.  \citet{bd79} revisited this reaction ten years
later treating the problem as one of scattering by a complex potential
(i.e., resonant scattering theory) but appear to have only accounted
for the ${\rm X}\ ^2\Sigma_u^+$ state.  All of these calculations are
in reasonable agreement with the laboratory results, to within the
quoted experimental factor of 2 uncertainty.

\citet{s89} calculated the rate coefficient for reaction~\ref{hm_ad},
following the quantum methodology of \citet{bd79}.  However, Sakimoto
paid attention to lower collision energies than considered by Bieniek
\& Dalgarno and also used new results for the ${\rm X}\ ^2\Sigma_u^+$
potential.  The resulting rate coefficient is a factor of almost 4
times larger than the measurement of \citet{sff67}.  Sakimoto writes
that this may be due to his having not taken into account the
contribution of the ${\rm B}\ ^2\Sigma_g^+$ state.  Sakimoto also
presents a classical Langevin limit rate coefficient which falls
outside of the experimental uncertainty limits.

Using resonant scattering theory and a local complex potential,
Launay, Le Dourneuf, \& Zeippen (1991) calculated the rate coefficient
for reaction~\ref{hm_ad} taking into account only the ${\rm X}\
^2\Sigma_u^+$ state.  Launay et~al.\ present results for their
potentials V1 and V2.  Each resulting rate coefficient lies within the
factor of 2 experimental error bars.  However, the energy dependence
calculated by Launay et~al.\ differs from that of \citet{bd69} above
$\approx 2,000 \: {\rm K}$.  Also, it appears that the results using
the potential V2 are incorrect as is discussed in more detail by 
\v{C}\'{\i}\v{z}ek, Hor\'{a}\v{c}ek, \& Domcke (1998).

The most recent calculation for reaction~\ref{hm_ad} is that of
\citet{chd98}.  They use a nonlocal resonance model and a new
potential for the ${\rm X}\ ^2\Sigma_u^+$ state.  They do not take
into account the ${\rm B}\ ^2\Sigma_g^+$ state.  The resulting rate
coefficient lies a factor of $\approx 3$ above the experimental
results of \citet{sff67}.  Their classical Langevin limit rate
coefficient is in good agreement with the experiment but lies a factor
of $\approx 2$ below the Langevin results of \citet{s89}.

Taking the lower limit for the experimental rate coefficient and the
highest calculated value, yields almost an order of magnitude spread.
This is surprisingly large for such a simple reaction.  A partial
explanation for the disagreement between the various theoretical
calculations may lie in the $\mHtm$ potentials used.  Amaya-Tapia,
Cisneros, \& Russek (1986) showed that theory had not yet converged to
a single set of potentials for either the ${\rm X}\ ^2\Sigma_u^+$ or
${\rm B}\ ^2\Sigma_g^+$ state of $\mHtm$.  The most recent
calculations for reaction~\ref{hm_ad} use differing potentials for the
${\rm X}\ ^2\Sigma_u^+$ state and have ignored the ${\rm B}\
^2\Sigma_g^+$ state \citep{s89,llz91,chd98} and all yield differing
results.  With theory unable to converge to the same value for the
rate coefficient, it appears that the only hope of improving our
understanding of reaction~\ref{hm_ad} lies with carrying out new
laboratory measurements.

\subsection{Mutual Neutralization}

The first published theoretical rate coefficient for
reaction~\ref{hm_mn} at cosmological temperatures appears to be the
Landau-Zener (LZ) results of \citet{bl55}.  Moseley, Aberth, \&
Peterson (1970) are the only group to have measured
reaction~\ref{hm_mn} at collision energies below 3~eV.  They carried
out measurements to energies as low as 0.1~eV.  Using their cross
section results and extrapolating below 0.1~eV, they produced an
experimentally-derived rate coefficient \citep{map70,pams71}.  Their
experimental rate coefficient is a factor of $\approx 3$ times larger 
than the LZ results of \citet{bl55}.

\citet{dm73} write that the rate coefficient for
\begin{equation}
\Hm + {\rm X}^+ \rightarrow \mH + {\rm X}  \label{xm_mn}
\end{equation}
is ``fairly insensitive to the nature of the ions ${\rm X}^+$''.  They
recommend a rate coefficient for all such reactions which is based on
the results of \citet{pams71} for reaction~\ref{hm_mn}.  Surprisingly,
the rate coefficient of Dalgarno \& McCray is a factor of $\approx 2$
times smaller than that of Peterson et~al.  This may be a partial
explanation as to why Dalgarno \& McCray estimate that their
recommended rate coefficient is accurate to ``perhaps a factor of 2''.
\citet{ph80}, in their commonly cited compilation of rate coefficients
for gas phase chemistry, use the recommended rate coefficient of
\citet{dm73}.  This then made its way into the recommended data of
\citet{dw84} and then into \citet{sk87}.

\citet{fk86} calculated reaction~\ref{hm_mn} using a quantum
close-coupling treatment.  Their cross section results are in good
agreement with the LZ calculations of \citet{bl55} but a factor of
$\approx 3$ times smaller than the measurements of \citet{map70}.
\citet{cdg99} used the cross section calculations of Fussen \& Kubach
to produce a rate coefficient.

\citet{dl87} present two rate coefficients for reaction~\ref{hm_mn}, a
high value and a low value.  The higher value is supposedly based on
the results of \citet{map70}, but as can be seen in Fig~\ref{mn}, lies
a factor of $\approx 4$ below their results.  The lower value is
supposed to be derived from the experimental results of \citet{sktb84}
and Peart, Bennett, \& Dolder (1985) which lie close to the
theoretical cross section predictions of \citet{bl55}.  However, these
experimental results were carried out for collision energies in excess
of 5~eV which is of little relevance for early Universe chemistry.
Also, Fig.~\ref{mn} shows that the lower recommended value of Dalgarno
\& Lepp is a factor of $\approx 3$ times smaller than that of
\citet{bl55}.  The cause for the differences for each rate coefficient
of Dalgarno \& Lepp and their stated sources is unclear.  Perhaps both
represent typographical errors.  The lower rate coefficient of
Dalgarno \& Lepp was adopted by \citet{aanz97a} for their primordial
chemistry.

Clearly, there is still a large uncertainty in the rate coefficient
for reaction~\ref{hm_mn}.  Is is true that LZ and quantum
close-coupling calculations are in good agreement for $T < 10^4 \:
{\rm K}$.  But the only measurement at the relevant collision energies
lies a factor of $\approx 3$ higher.  Cross section measurements at
collision energies above 3~eV have been carried out by a number of
groups \citep{sktb84,pbd85,ph92} and these are in reasonable agreement
with the calculations of \citet{bl55} and \citet{fk86}.  However,
since the work of \citet{map70} no further measurements have been
carried out at collision energies relevant for early Universe
chemistry.  Additionally, no explanation has been given to date as to
why the data of \citet{map70} may be wrong
\citep[cf.,][]{sktb84,pbd85,ph92}.  It is clear that in order to
resolve this issue, new laboratory measurements are needed for
reaction~\ref{hm_mn} at collision energies below 1~eV.

\section{Computational method}
\label{code}

To aid us in assessing the impact of the existing uncertainties in the
associative detachment and mutual neutralization rate coefficients on
the cooling and collapse of primordial gas, we have performed a number
of simulations of the collapse of gas into protogalactic halos using 
a numerical technique known as smoothed particle hydrodynamics, or
SPH \citep{gm77,lucy77,mon92}. SPH is a Lagrangian method for simulating 
astrophysical fluid flows, in which the fluid is represented by an ensemble 
of `particles', with flow quantities at a particular point obtained by 
averaging over an appropriate subset of neighbouring SPH particles.
It should be noted that the particles used in SPH simulations in no 
sense correspond to actual gas particles; in typical astrophysical 
simulations, the former have masses greatly in excess of a solar mass
and so represent more than $10^{57}$ actual gas particles. Instead,
each SPH particle simply corresponds to a small region of the flow,
with fixed mass but variable volume \citep[cf., Eulerian grid-based codes,
such as that described in][where the volume of each grid 
cell is fixed, but the mass of fluid contained within it varies]{sn92}.
To perform our simulations, we used a a modified version of the Gadget SPH
code \citep{syw01}. Full details of our modifications are given elsewhere 
\citep{jgk05}, but we briefly summarize them here.

The most significant modification that we have made to the code is the
addition of the necessary framework for following the non-equilibrium
chemistry of $\mHt$ in the protogalactic gas. To incorporate this into
the code, we associate a set of chemical abundances with each SPH 
particle. Just as with the other fluid properties, such as density or 
internal energy, these abundances represent averages over the local 
fluid flow. For each SPH particle, we therefore must solve a set of
chemical equations of the form
\begin{equation}
\frac{{\rm d}\,x_{i}}{{\rm d}\, t} = C_{i} - D_{i} x_{i}
\end{equation}
where $x_{i}$ is the fractional abundance of species $i$, computed with
respect to the total number density of hydrogen nuclei (i.e.,
$x_{i} = n_{i} / n$, where $n_{i}$ is the number density of species $i$
and $n$ is the number density of hydrogen nuclei), and where $C_{i}$ and 
$D_{i}$ are terms representing the chemical creation and destruction of 
species $i$. The values of the creation and destruction terms generally 
depend on both the density and temperature of the gas, as well as on its 
chemical composition.

In our code, we follow the chemistry of six species: $\me$,
$\Hp$, $\mH$, $\Hm$, $\mHtp$, and $\mHt$. Two of these species, $\Hm$
and $\mHtp$, reach chemical equilibrium on very short timescales.
In the case of $\Hm$, one can readily show that equilibrium is reached 
on a timescale
\begin{equation}
t_{\rm eq, \Hm} \simeq \left(k_{\rm ad} n_{\mH} + k_{\rm mn} n_{\Hp}
\right)^{-1},
\end{equation}
where $n_{\mH}$ is the hydrogen atom density, $n_{\Hp}$ is the $\Hp$
density, and $k_{\rm ad}$ and $k_{\rm mn}$ are the associative
detachment and mutual neutralization rate coefficients, respectively.
We assume that other processes responsible for destroying $\Hm$, such
as photodetachment (reaction 6) can be neglected. If $n_{\Hp} \ll
(k_{\rm ad} / k_{\rm mn}) n_{\mH}$, then the associative detachment
term dominates and since $k_{\rm ad} \sim 10^{-9} \: {\rm cm^{3}} \:
{\rm s^{-1}}$ (to within an order of magnitude), it follows that
$t_{\rm eq, \Hm} \sim 10^{9} n_{\mH}^{-1} \: {\rm s}$. On the other
hand, if $n_{\Hp} \gg (k_{\rm ad} / k_{\rm mn}) n_{\mH}$, then the
mutual neutralization term dominates, in which case $t_{\rm eq, \Hm}
\sim 5 \times 10^{5} T^{1/2} n_{\Hp}^{-1} \: {\rm s}$, again to within
an order of magnitude. In either case, the equilibrium timescale is
very short. The largest value we find for it occurs within the neutral, 
low density gas present at the beginning of runs A1-A9 (see 
Section~\ref{cold-init} below), but even here, where 
$n_{\mH} \sim 10^{-3}$, its value is only $t_{\rm eq, \Hm} \sim 10^{12} 
\: {\rm s}$, which is significantly smaller than either the cooling time 
and the free-fall time of the gas (which are both typically much longer 
than a ${\rm Myr}$, as we demonstrate in Section~\ref{hot_uv}). In more
ionized gas, or at higher densities, $t_{\rm eq, \Hm}$ becomes even 
smaller relative to the cooling time $t_{\rm cool}$ and the free-fall
time $t_{\rm ff}$.

The case of $\mHtp$ is very similar: one can show that it reaches 
chemical equilibrium on a timescale
\begin{equation}
t_{\rm eq, \mHtp} \simeq \left(k_{\rm ct} n_{\mH} + k_{\rm dr} n_{\me} 
\right)^{-1},
\end{equation}
where $k_{\rm ct}$ and $k_{\rm dr}$ are the rate coefficients of the
charge transfer and dissociative recombination reactions which are the
main processes responsible for removing $\mHtp$ from the gas. (For
reference, these are reactions 4 and 7 in
Table~\ref{chemtab}). Evaluating this expression, we obtain $t_{\rm
eq, \mHtp} \sim 1.6 \times 10^{9} n_{\mH}^{-1} \: {\rm s}$ in the
low ionization limit, or $t_{\rm eq, \mHtp} \sim 10^{8} n_{\me}^{-1} \:
{\rm s}$ in the high ionization limit. Again, these timescales
are both very much shorter than either the cooling time or the
free-fall time of the gas.

Because $\Hm$ and $\mHtp$ reach equilibrium so quickly, we do not
attempt to follow their chemical equilibrium directly in our
simulations. Instead, their abundances are computed only as required,
under the assumption that both reach equilibrium instantaneously. 
This approximation introduces a certain amount of error into the
computed $\mHt$ abundance. However, provided that the timesteps used 
to evolve the SPH particles in the simulation are long compared 
compared to $t_{\rm eq, \Hm}$ and $t_{\rm eq, \mHtp}$, which we have
verified is the case for our simulations, the size of this error will
be negligible compared to that arising from the rate coefficient
uncertainties that are the subject of this paper. Having made this
approximation, we are left with four non-equilibrium species, and thus
in theory four chemical equations. However, since charge conservation
implies that
\begin{equation}
 x_{\Hp} + x_{\mHtp} = x_{\me} + x_{\Hm}, \label{cons1}
\end{equation}
while conservation of the total number of hydrogen nuclei implies that
\begin{equation}
 x_{\mH} + x_{\Hp} + x_{\Hm} + 2x_{\mHtp} + 2x_{\mHt} = 1, \label{cons2}
\end{equation}
we can reduce the actual number of equations that we must solve to only two. 
We choose to solve for $x_{\mHt}$ and $x_{\Hp}$, obtaining $x_{\me}$ and
$x_{\mH}$ from Equations~\ref{cons1}-\ref{cons2}, but other choices are of 
course possible.  Our choice here should make no difference to our final 
results.

To solve the chemical equations for a given SPH particle, we make use
of a technique known as operator splitting. We assume that within the
current timestep of the SPH particle, we can solve for the density
evolution of the gas separately from its chemical evolution. The
density evolution can then be computed using the same prescription as
in standard SPH (see Springel et al.\ 2001 for details), and the
updated gas density is then available for use in the chemical
equations. These are then solved implicitly using DVODE, a freely
available and well tested double precision ordinary differential
equation solver \citep{bbh89}. Operator splitting introduces a certain
amount of error, as in practice the density should vary during the
chemical timestep. However, the SPH algorithm naturally limits the
extent to which the density can change during a single SPH particle
timestep, by making particles take shorter timesteps in rapidly
evolving regions. We therefore expect the error introduced by this
technique to be small as is discussed further in Section~\ref{init}.

In common with other authors, we use a simplified reaction network
that does not include the chemistry of minor coolants such as ${\rm
HD}$ or ${\rm LiH}$.  We also neglect any effects due to helium
chemistry. Neglect of the minor coolants is justified by the fact that
$\mHt$ dominates the cooling of the gas for the all of the
temperatures and densities found in our simulations \citep{fpf01}.  At
the worst, we may overestimate the final temperature of the gas
slightly. Neglect of the helium chemistry is also easily justified,
provided that we assume that the bulk of the helium in the gas is in
neutral form, as in this case the only reactions involving helium that
play any role in determining the $\mHt$ abundance -- the collisional
dissociation of $\Hm$ and $\mHt$ by ${\rm He}$ -- are far less
effective than the corresponding reactions with $\mH$ \citep{aanz97a}.
So the error in the final $\mHt$ abundance will be small. If
significant amounts of ionized helium are present, then our assumption
introduces a larger error, since we will underestimate the actual
electron abundance in the gas, and hence the $\Hm$ formation
rate. However, even in this case, we would expect the error in the
final $\mHt$ abundance to be relatively small, owing to the small
abundance of helium relative to hydrogen. The chemical reactions
included in our network are summarized in Table~\ref{chemtab}.  In most
cases, we also list the source used for the adopted rate
coefficient. The exceptions are the associative detachment and mutual
neutralization reactions, which we discuss further below.

Table~\ref{chemtab} lists three photochemical reactions: the
photodetachment of $\Hm$ and the photodissociation of $\mHtp$ and
$\mHt$. To compute the rates for these reactions, we assumed that
our simulated protogalaxies were illuminated by an external background 
radiation spectrum with the shape of a $10^{5} \: {\rm K}$ black-body, 
as should be typical of the brightest population III stars \citep{coj00}.  
We cut off this spectrum at energies greater than
$13.6 \: {\rm eV}$ to account for absorption by neutral hydrogen in
the protogalactic gas and in the intergalactic medium. The strength of
the background is specified in terms of the flux at the Lyman limit,
$J(\nu_{\alpha}) = 10^{-21} J_{21} \: {\rm erg} \: {\rm s^{-1}} \:
{\rm cm^{-2}} \: {\rm Hz^{-1}} \: {\rm sr^{-1}}$. 

If sufficient $\mHt$ forms within the protogalaxy, it will begin to
self-shield, reducing the effective photodissociation rate. An exact
treatment of the effects of self-shielding is computationally
unfeasible, as it would require us to solve for the full spatial,
angular and frequency dependence of the radiation field at every
timestep. Instead, we have chosen to incorporate it in an approximate
manner. We assume that the dominant contribution to the self-shielding
at a given point in the protogalaxy comes from gas close to that
point, and so only include the contribution to the self-shielding that
comes from the nearby $\mHt$. To implement this approximation
numerically, we make use of the fact that Gadget already defines a
suitable local length scale: the SPH smoothing length $h$, which
characterizes the scale over which the flow variables are averaged. In
Gadget, as in all modern SPH codes, $h$ is allowed to vary from point
to point within the flow and is automatically adjusted in order to
keep the mass enclosed within a sphere of radius $h$ roughly
constant. Further details can be found in Springel et al.\ (2001).  In
our calculation of the $\mHt$ column density used to compute the
degree of self-shielding we include only $\mHt$ which lies within a
single smoothing length of the point of interest. We justify this
approximation by noting that in our simulations, widely separated SPH
particles typically move with a significant velocity relative to one
another. Consequently, the contribution to the total absorption
arising from one particle is Doppler shifted when viewed from the rest
frame of the other particle. If this Doppler shift is large compared
to the line widths of the Lyman-Werner band transitions, then the
effect is to dramatically reduce the extent to which the absorption
contributes to the total self-shielding.  On the other hand, gas close
to the point of interest will typically have a much smaller relative
velocity, and so will contribute far more effectively.  Our
approximation considers only the latter contribution, and assumes that
the former contribution is completely negligible. In practice, of
course, the distant gas is likely to contribute to some non-negligible
extent, and so we will tend to underestimate the true amount of
self-shielding, unless the gas infall is highly supersonic.
Nevertheless, we believe that our approximation remains useful as it
is computationally efficient, and also represents an improvement over
previous optically thin treatments \citep[e.g.,][]{mba01,rgs02}.

Finally, we assume that ionization from X-rays or cosmic rays is
negligible, although previous work suggests that even if a low level
of ionization is present, it will not have a major effect on the
outcome of the collapse (Glover \& Brand 2003; Machacek et al.\ 2003).

A second major modification that we have made to the Gadget code is
the inclusion of a treatment of radiative heating and cooling. Cooling
in our model comes from three main sources: electron impact excitation
of atomic hydrogen (a.k.a.\ Lyman-$\alpha$ cooling), which is
effective only above about $8000 \: {\rm K}$, rotational and
vibrational excitation of $\mHt$, and Compton
cooling. Rates for Lyman-$\alpha$ cooling and Compton cooling were
taken from \citet{cen92}, while for $\mHt$ rovibrational cooling we
used a cooling function from \citet{lpf99}. In models where an
ultraviolet background is present, we include the effects of heating
from the photodissociation of $\mHt$, assuming that $0.4 \: {\rm eV}$
of energy per photodissociation is deposited as heat \citep{bd77}. We
also include heating due to the ultraviolet pumping of $\mHt$,
following \citet{bht90}, although this is only important in high
density gas ($n \geq 10^{4} \: {\rm cm^{-3}}$).

To incorporate the radiative heating and cooling terms into Gadget, we
again use an operator splitting technique.  In this case, we assume 
that the change in the internal energy of the gas due to pressure work
can be computed separately from that due to radiative heating and 
cooling. The former can then be calculated in the same fashion as in 
the standard Gadget code, while the latter can be computed by solving:
\begin{equation}
\frac{{\rm d}\,\epsilon}{{\rm d}\, t} = \Gamma - \Lambda,
\end{equation}
where $\epsilon$ is our initial estimate of the internal energy density of 
the gas, which already includes the effects of the pressure work term,
$\Gamma$ is the radiative heating rate per unit volume and $\Lambda$ is the 
radiative cooling rate per unit volume. We solve this equation implicitly 
using DVODE at the same time that we solve the chemical equations.
Again, the use of operator splitting introduces some error into the thermal
evolution of the gas, but, as before, we expect this error to be small
(see Section~\ref{init} for further discussion).

Finally, to allow us to represent gas that has collapsed beyond the
resolution limit of the simulation in a numerically robust manner, we
have modified the code to allow it to create sink particles --
massive, non-gaseous particles, designed to represent dense cores,
that can accrete gas from their surroundings but otherwise interact
only via gravity \citep{bbp95}. The design and implementation of our
sink particle algorithm is discussed elsewhere \citep{jap05}.

\section{Initial conditions}
\label{init}

As we wish to be able to run a large number of simulations of
protogalactic collapse, we have chosen to limit the computational cost
of each simulation by starting from somewhat simplified initial
conditions. Since we are not particularly interested (in this paper at
least) in following the assembly history of the dark matter halo in
which the protogalaxy sits, or in studying the response of the halo to
the cooling of the gas, we chose to model the influence of the halo by
using a fixed background potential. To construct this potential, we
assumed that the halo was spherically symmetric, with the density
profile of a Burkert halo \citep{burk95}:
\begin{equation}
\rho_{\rm dm}(r) = \frac{\rho_{\rm dm, 0}}{1 + (r^2 / r_{\rm c}^2)}.
\end{equation}
We took the central density of the halo to be $\rho_{\rm dm, 0} = 0.13
\: {\rm M_{\odot}} \: {\rm pc}^{-3}$ and also specified the total mass
of dark matter in the halo, $M_{\rm dm} = 10^{7} \: {\rm
M_{\odot}}$. The core radius $r_{\rm c}$ is calculated internally by
Gadget, based on $M_{\rm dm}$ and the initial redshift $z$. For the
gas, we assumed an initially uniform distribution, with an initial
density $\rho_{\rm g}$, taken to be equal to the cosmological
background density. The initial temperature of the gas was also taken
to be uniform, with a value $T_{\rm g}$, the choice of which is
discussed below.

The quantity of gas present in our simulations was taken to be a
fraction $\Omega_{\rm b} / \Omega_{\rm dm}$ of the total mass of dark
matter, where $\Omega_{\rm dm} = \Omega_{\rm m} - \Omega_{\rm b}$. We
took values for the cosmological parameters from \citet{wmap}, and so
$\Omega_{\rm b} = 0.047$ and $\Omega_{\rm m} = 0.29$, giving us a
total gas mass of $M_{\rm g} = 0.047 / (0.29 - 0.047) \times 10^{7} \:
{\rm M_{\odot}} = 1.934 \times 10^{6} \: {\rm M_{\odot}}$. In most of
our simulations, we used $32768$ SPH particles to represent this gas,
giving each SPH particle a mass $M_{\rm part} \simeq 59 \: {\rm
M}_{\odot}$. In order to properly resolve gravitationally bound clumps
(or other gravitationally bound structures) in SPH simulations, they
must be represented by at least twice as many SPH particles as are
used in the SPH smoothing kernel \citep{bb97}. In our simulations,
our smoothing kernel encompasses approximately 40 particles -- for 
reasons of computational efficiency, the number is allowed to vary 
slightly, but never by more than 5 particles -- and so our minimum mass 
resolution is $M_{\rm res} \simeq 80 M_{\rm part} \simeq 4720 \: 
{\rm M_{\odot}}$. To verify that our results are insensitive to the value of 
$M_{\rm res}$, we have repeated several of our simulations using a 
larger number of SPH particles, $131072$, corresponding to a
four times smaller value of $M_{\rm res}$. We find only minor 
differences (of the order of 10\% or less) between the temperature and
density evolution in these runs and in the lower resolution runs 
discussed below. Since the SPH particle timesteps in these higher 
resolution runs will typically be shorter than those in our lower
resolution runs, the size of the error introduced by our use of 
operator splitting will also be different. Therefore, the close 
agreement of the results of these runs with their lower resolution
counterparts also helps to reassure us that the splitting errors in 
both sets of runs are small.

To prevent artificial fragmentation or other numerical artifacts from
affecting our results, it is necessary either to halt the simulation
before the local Jeans mass, $M_{\rm J}$ falls below $M_{\rm res}$ in
any part of the simulation volume, or to use sink particles to
represent regions where $M_{\rm J} < M_{\rm res}$. We have chosen the
latter course, and so create sinks in regions where the gas density
exceeds $6 \: {\rm M_{\odot}} \: {\rm pc}^{-3}$ using the algorithm
described in \citet{jap05}.  This corresponds to a hydrogen atom
number density $n_{\rm crit} \sim 200 \: {\rm cm^{-3}}$.  The Jeans
mass, which is defined here as\footnote{Note that in this definition,
we assume that the self-gravity of the gas dominates on small scales
within the halo, since the gas is dissipative and so can collapse to
much higher densities than the non-dissipative dark matter.}
\begin{equation}
M_{\rm J} = \frac{\pi^{5/2}}{6} G^{-3/2} \rho_{\rm g}^{-1/2} 
c_{\rm s}^{3},
\end{equation}
where $\rho_{\rm g}$ is the gas density and $c_{\rm s}$ is the adiabatic 
sound speed, has a value at this number density of 
\begin{equation}
 M_{\rm J} \simeq 5000 \left(\frac{T}{100 \: {\rm K}}\right)^{3/2} \;
 {\rm M_{\odot}}.
\end{equation}
 This is greater than $M_{\rm res}$ provided that the temperature of
the dense gas exceeds $100 \: {\rm K}$. As we shall see later, the
minimum gas temperature reached by dense, gravitationally collapsed 
gas in our simulations is typically no smaller than $150 \: {\rm K}$ and 
so our simulations remain well-resolved up to the point at which sink 
particles are created.

Once created, sink particles can accrete gas from their surroundings,
provided that the SPH particles representing the gas come within a
specified radius of the sink particle (the accretion radius $r_{\rm
acc}$), are gravitationally bound to the sink, and satisfy certain 
other conditions (see \citealt{jap05} for details). In our simulations, we 
set $r_{\rm acc} = 20 \: {\rm pc}$.  This value is chosen to be somewhat
larger than Jeans length of the gas at $n_{\rm crit}$, but is admittedly
somewhat arbitrary. However, since SPH particles within $r_{\rm acc}$ 
will only be accreted if they satisfy {\em all} of the necessary criteria,
the outcome of our simulations should be insensitive to the precise value 
chosen for $r_{\rm acc}$.

We initialized each of our simulations at a redshift $z = 20$ and
allowed them to run for $220 \: {\rm Myr}$; given our adopted
cosmological parameters, this interval corresponds to approximately
1.25 Hubble times, with the simulations terminating at a redshift $z
\simeq 11.2$. Protogalaxies which fail to cool and collapse during
this interval are unlikely to get the chance to do so thereafter, as
the typical interval between major mergers of dark matter halos is of
the order of a Hubble time \citep{lc93}.

Finally, it should be noted that the initial conditions described here
are undoubtedly highly simplified and are unlikely to properly
represent the full hydrodynamics of protogalactic formation. However,
they should be more than sufficient for our purposes, as we are
primarily interested in the {\em difference} in the outcomes of
simulations run using different values for the mutual neutralization
and associative detachment rate coefficients, and in this respect we
would expect the behaviour of these simple models to be a reliable
guide to the behaviour of more complex, but ultimately more realistic
models.

\section{Results}
\label{result}

To investigate how sensitive the cooling and collapse of protogalactic
gas are to the choice of mutual neutralization and/or associative
detachment rate coefficients, we have performed five sets of
simulations, with initial conditions as summarized in the previous
section and in Table~\ref{runtab}. For each set of parameters listed
in Table~\ref{runtab}, we performed nine separate runs in which the
mutual neutralization and associative detachment rate coefficients
were varied individually.  For runs 1-3, we adopted the lower value
for the mutual neutralization rate coefficient from \citet{dl87},
while for the associative detachment rate coefficient we used values
of $0.65 \times 10^{-9}$, $1.30 \times 10^{-9}$ and $5.00 \times
10^{-9} \: {\rm cm^{3}}~{\rm s}^{-1}$ respectively, where the central
value corresponds to the value measured by \citet{sff67} and the other
values represent lower and upper bounds on plausible values, as
previously discussed in Section~\ref{rates}. For runs 4-6, we adopted
the mutual neutralization rate coefficient of \citet{cdg99} and varied
the associative detachment data as before. Finally, for runs 7-9 we
repeated this procedure using the mutual neutralization rate
coefficient of \citet{pams71}. For ease of reference, the nine
different combinations of rate coefficients used are listed in
Table~\ref{settab}.

\subsection{Cold initial conditions}
\label{cold-init}

Our first set of simulations (runs A1-A9) examined collapse beginning
from cold initial conditions. In these runs we took initial values for
the fractional ionization and $\mHt$ abundance from the IGM chemistry
model of Stancil et al.\ (1998), and adopted an initial gas
temperature $T_{\rm g} = 12 \: {\rm K}$, corresponding to the value of
the IGM temperature at $z=20$ in the absence of any form of
preheating.

\begin{figure}[Ht]
\centering
\epsfig{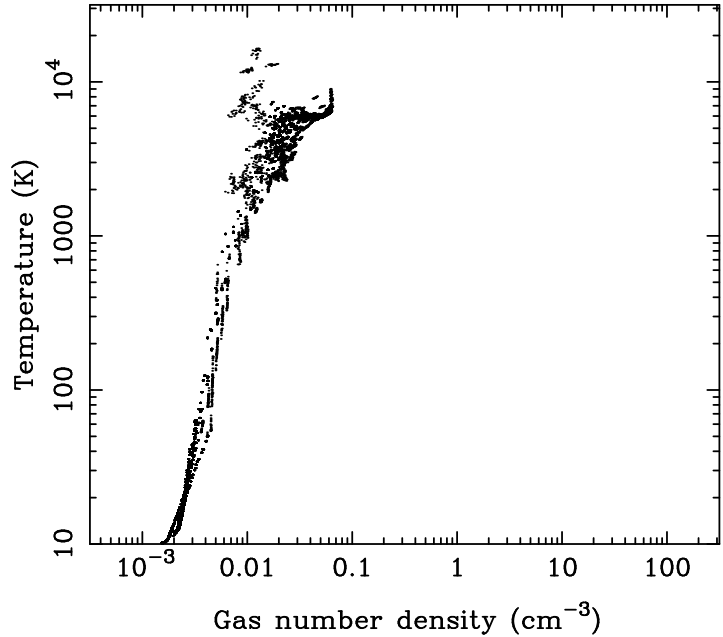}
\caption{The density and temperature distribution at the end of run
A1. Each point in the plot represents the gas density and temperature
associated with an individual SPH particle.}
\label{nTcold}
\end{figure}

The state of the gas at the end of run A1 is illustrated by the plot
of temperature versus number density shown in
Figure~\ref{nTcold}. Infalling cold gas has been shock-heated to $T >
10^{4} \: {\rm K}$ within the potential well of the dark matter halo,
and some cooling of the highest density gas is evident, although this
cooling is not yet far advanced. By the end of the simulation, the 
density of the central gas has reached $n_{\rm c} \simeq 6.5 
\times 10^{-2} \: {\rm cm}^{-3}$ and its temperature is $T_{\rm c} 
\simeq 6900 \: {\rm K}$ (although note that because of the nature of
the SPH algorithm, these values actually represent a weighted average
over gas within a single smoothing length of the center of the halo).
Its fractional ionization remains fairly low, 
$x_{\rm \me, c} \simeq 2.0 \times 10^{-4}$, as does its
$\mHt$ abundance, $x_{\rm \mHt, c} = 1.8 \times 10^{-5}$, though the
latter has increased by nearly an order of magnitude over its initial
value. The cooling time of this gas is long -- approximately $570 \:
{\rm Myr}$ -- although it will decrease significantly once more $\mHt$
forms.

The other runs in this set of simulations (runs A2-A9) give very
similar results.  In Table~\ref{run_a_results}, we list the central
density, temperature, fractional ionization and $\mHt$ abundance at
the end of each of these runs. From the table it is clear that while
changes in the mutual neutralization rate coefficient have only a very
minor effect on the final state of the gas, changes in the associative
detachment rate coefficient have a far more noticeable effect -- in
runs with a high value of the associative detachment rate coefficient,
more $\mHt$ is produced, and the extra cooling that this provides
leads to a lower central temperature and consequently a higher central
density at the end of the run. For variations of the associative
detachment rate coefficient within the range that we consider to be
plausible, we find a variation of approximately a factor of three in
the final $\mHt$ abundance. However, the variation in the final
temperature and density is considerably smaller, approximately 10\%,
since in none of the models has enough $\mHt$ formed to provide
particularly effective cooling.

The fact that the results of these runs are insensitive to the value
of the mutual neutralization rate coefficient is easy to understand if
one considers that at a fractional ionization of order $10^{-4}$, the
net rate of destruction of $\Hm$ by mutual neutralization is two to
four orders of magnitude smaller than the rate of destruction by
associative detachment, meaning that the latter always dominates.  The
sensitivity of the results to the choice of associative detachment
rate coefficient comes about in this case due to the influence of a
third process, the collisional dissociation of $\Hm$ by atomic
hydrogen (reaction 15), which has a rate coefficient which is
competitive with that of associative detachment at high
temperature. However, by the end of the simulations, the collisional
dissociation rate in the central gas has already become smaller than
the associative detachment rate, and so we would expect its influence
on future $\mHt$ formation to be slight.

Ultimately, while the differences we find between the results of the
various runs are interesting, they do not appear to be particularly
important, since in every case the outcome of the simulation is
essentially the same -- the protogalactic gas fails to cool within a
Hubble time, and so is unlikely to have sufficient time to cool or
collapse before the protogalaxy merges with a larger structure.

The fact that gas in these runs fails to cool effectively is not
entirely unexpected, given our choice of halo mass and initial
redshift.  Previous work on protogalaxy formation in the cold IGM by
\citet{teg97} and others suggests that only gas in halos more massive
than some critical mass $M_{\rm crit}$ will cool effectively (reviewed
recently by \citealt{bl04}, \citealt{cf05}, and \citealt{g05}). For gas
collapsing at $z < 20$, \citet{teg97} find $M_{\rm crit} \gtrsim
10^{7} \: {\rm M_{\odot}}$, suggesting that cooling in our $10^{7} \:
{\rm M_{\odot}}$ halos will only be marginally effective. We therefore
performed two additional sets of simulations, using similar initial
conditions to runs A1-A9, but starting at $z = 30$ and $z = 40$.  
Appropriate adjustments were of course made to the initial density and
temperature of the cold gas.  We find more evidence for cooling in
these runs, in line with the theoretical expectations, but variations
in the associative detachment and mutual neutralization rate
coefficients continue to have no more than a small effect, with the
largest variation in the central gas density being of order 5\% and
the largest variation in the central $\mHt$ abundance being of order
15\%. We therefore conclude that our $z=20$ runs are giving us an
accurate picture of the influence of the uncertainties in the rate
coefficients in this particular scenario.

\subsection{Hot initial conditions}

Since the uncertainties in the rate coefficients do not appear to have
a large impact on the outcome of protogalactic collapse that begins
from cold initial conditions, we chose in our next set of runs (B1-B9)
to examine an alternative situation, in which we might expect the
uncertainties to have a greater effect. We took as initial conditions
gas which was hot ($T_{\rm g} = 10^{4} \: {\rm K}$) and fully ionized
($x_{\me} = 1.0$, $x_{\mHt} = 0.0$). The physical situation that these
initial conditions are intended to represent is that of a protogalaxy
forming within what \citet{oh03} term a `fossil' \hii region -- an
\hii region surrounding an ionizing source which has just switched
off, but the surrounding gas has not yet had time to cool and
recombine. Prior to cosmological reionization, such regions should be
relatively common, since the characteristic lifetimes of the likely
ionizing sources -- massive population III stars and/or active
galactic nuclei -- are significantly shorter than the Hubble time.

\begin{figure}[Ht]
\centering
\epsfig{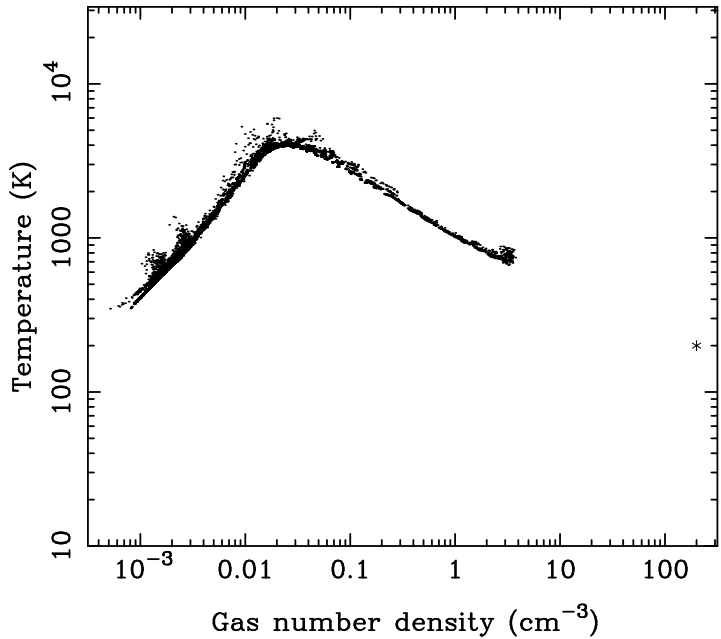}
\caption{The density and temperature distribution at the end of run
B1. Each point in the plot represents the gas density and temperature
associated with an individual SPH particle. The asterisk represents the 
nominal temperature and density of the gas comprising the sole sink 
particle immediately before the formation of the sink.}
\label{nT}
\end{figure}

In Figure~\ref{nT}, we show the state of the gas at the end of run
B1. The asterisk on the right hand side of the plot represents a sink
particle of mass $M_{\rm sink} \simeq 1.2 \times 10^{5} \: {\rm
M_{\odot}}$. Since we have no information on the true temperature and
density distribution of the gas represented by the sink particle,
beyond the knowledge that its density is greater than $200 \: {\rm
cm^{-3}}$, we cannot assigned the material within the sink particle to
its proper location in this plot. We have therefore assigned it a
nominal density of $200 \: {\rm cm^{-3}}$ and temperature of $200 \:
{\rm K}$, as these values are broadly representative of the properties
of the gas immediately prior to sink particle formation. The contrast
between this figure and Figure~\ref{nTcold} is striking. Far more high
density gas is present, especially if one takes the material in the
sink particle into account, and the temperature of the gas is much
lower: the minimum temperature reached by the end of the simulation in
gas which has not become part of the sink is $347 \: {\rm K}$, while
the lowest temperature reached in any of the gas during the course of
the run is $172 \: {\rm K}$. Note
that the reason that the latter value is lower is that we cease
tracking the temperature of gas once it becomes part of a sink
particle; consequently, the minimum temperature appears to increase
after the creation of the sink particle. Compared to these values, the
minimum temperature reached in run A1, which is $6940 \: {\rm K}$, is
a factor of twenty to fourty times larger.

\begin{figure}[Ht]
\centering
\epsfig{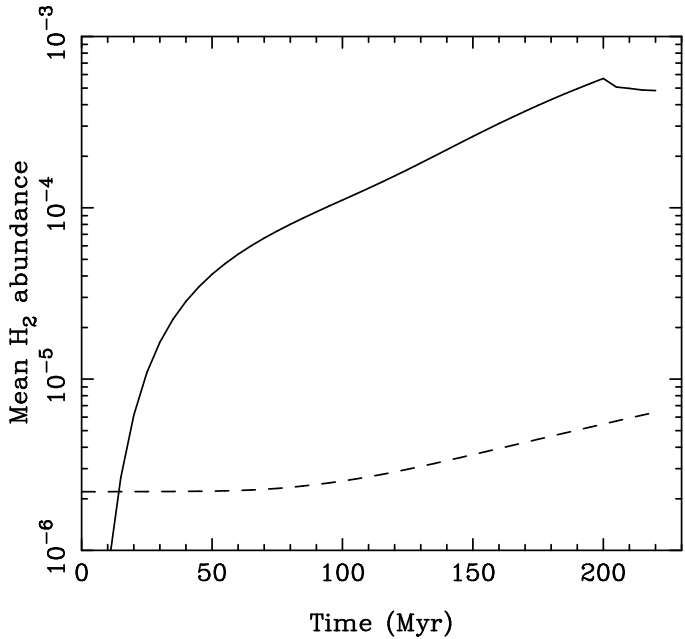}
\caption{The evolution with time of the mass-weighted mean $\mHt$ 
abundance in run A1 (represented by the dashed line) and run B1 
(represented by the solid line). The apparent decline in the latter at 
$t = 200 \: {\rm Myr}$ is due to the formation of a sink particle, as 
explained in the text.}
\label{H2B1}
\end{figure}

The reason for this dramatic difference becomes clear if one compares
the growth of the mass-weighted mean $\mHt$ fraction in runs A1 and 
B1. This quantity is given by
\begin{equation}
 x_{\rm \mHt, mean} = \frac{\Sigma_{i} x_{\mHt, i}}{N_{\rm sph}}
\end{equation}
where $x_{\mHt, i}$ is the fractional $\mHt$ abundance of the $i$-th
SPH particle and we sum over all $N_{\rm sph}$ SPH particles.  Since
all of the SPH particles have the same mass, this gives us a
mass-weighted value. The evolution of $x_{\rm \mHt, mean}$ during the 
two runs is plotted in Figure~\ref{H2B1}. Although run A1 has a higher
initial $\mHt$ abundance than run B1, it is clear that $\mHt$ forms
far more rapidly in the latter run, due to the high initial fractional
ionization, and it very quickly overtakes run A1. Similar differences
are seen if instead of comparing the mean $\mHt$ fraction, we compare
the evolution of the $\mHt$ fraction at the center of the density 
distribution, $x_{\rm \mHt, c}$. Note also that the apparent
decline in $x_{\rm \mHt, mean}$ at $t \geq 200 \: {\rm Myr}$ is simply
an artifact of the creation of the sink particle, since we cease
tracking the $\mHt$ content of gas that becomes part of a sink
particle. The much higher $\mHt$ fraction found in run B1 allows the
gas to cool efficiently to a few hundred K, and consequently enables
it to collapse to high density. It is reasonable to suppose that some
fraction of this collapsed gas will go on to form stars.

\begin{figure}
\centering
\epsfig{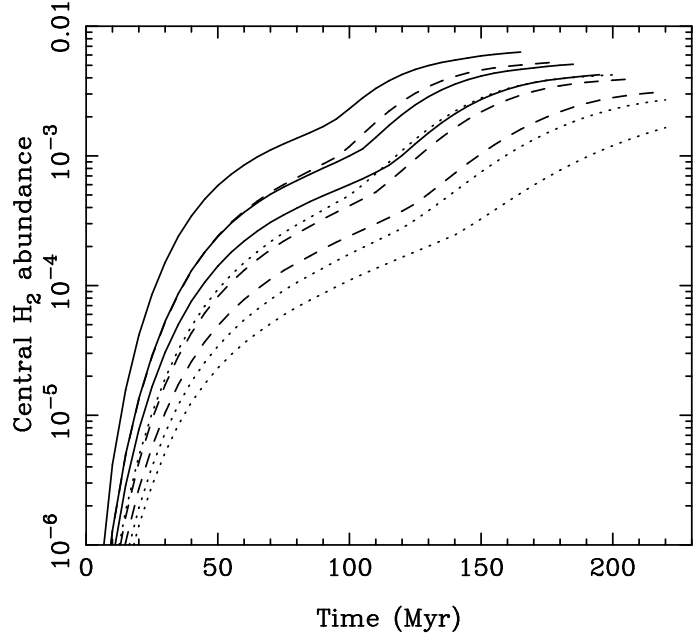}
\caption{The evolution with time of the central $\mHt$ abundance
found in the gas in runs B1-B9.  From bottom to top, the dotted lines
correspond to runs B7, B8, and B9 respectively, the dashed lines to
runs B4, B5, and B6 and the solid lines to runs B1, B2, and B3.  For
clarity, we only plot the evolution up until the point at which a sink
particle forms (or until the end of the run, if no sink forms).}
\label{maxH2-B}
\end{figure}

\begin{figure}
\centering
\epsfig{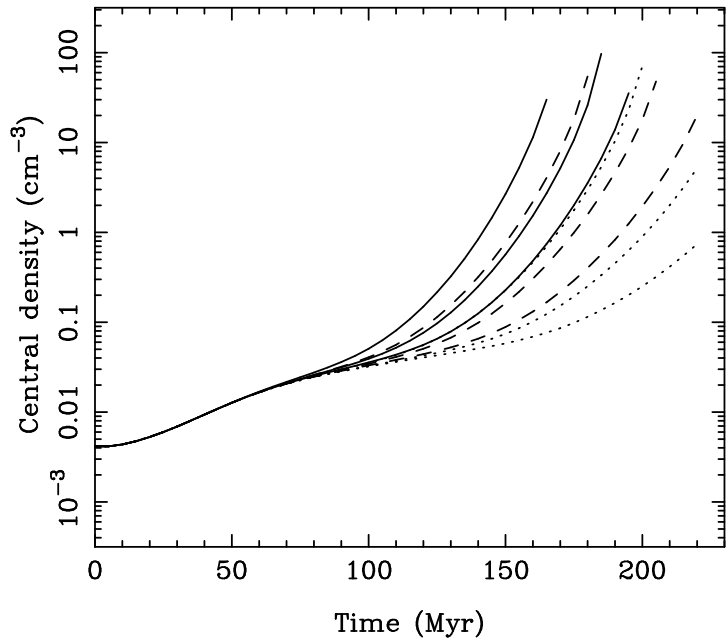}
\caption{Same as Figure~\ref{maxH2-B}, but for the central density of the 
gas.}
\label{maxn-B}
\end{figure}

\begin{figure}
\centering
\epsfig{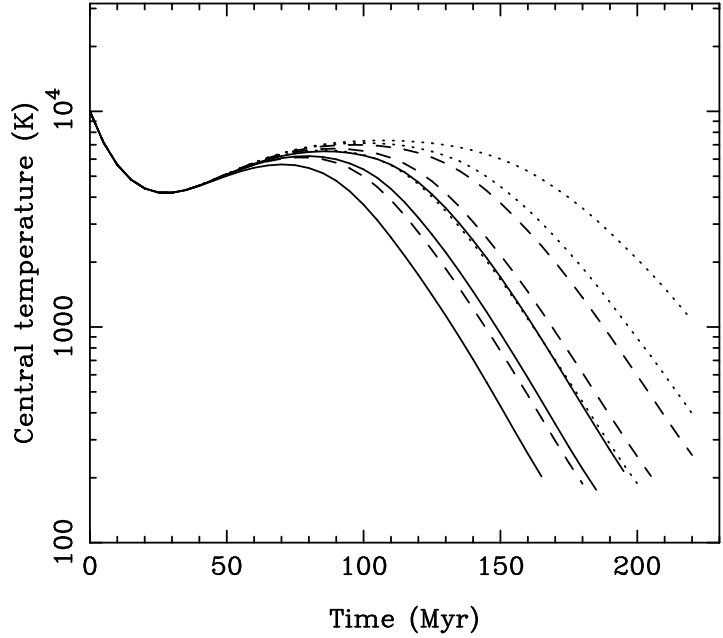}
\caption{Same as Figure~\ref{maxH2-B}, but for the central
temperature of the gas. Note that from bottom to top, the solid lines
now correspond to runs B3, B2, and B1 respectively, the dashed lines
to runs B6, B5, and B4 and the dotted lines to runs B9, B8, and B7.}
\label{minT-B}
\end{figure}

\begin{figure}
\centering
\epsfig{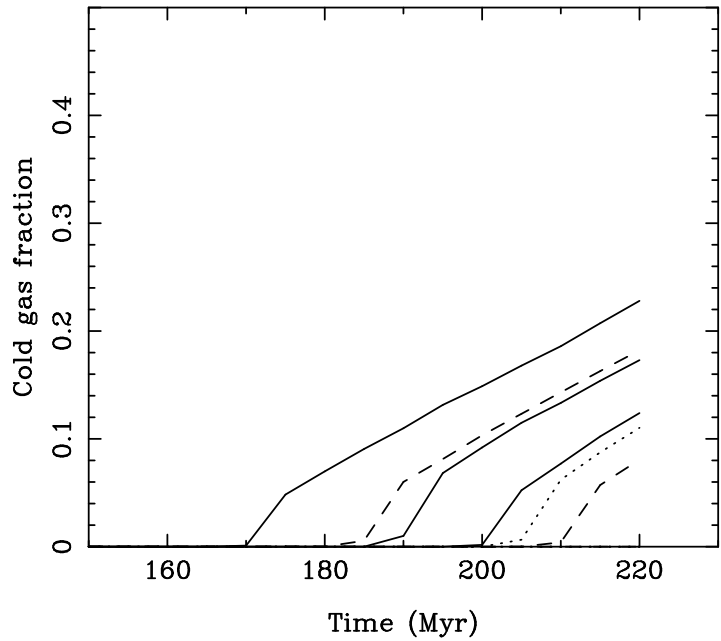}
\caption{The evolution with time of $f_{\rm cc}$, the fraction of 
cool, condensed gas.  The
labeling of the lines is as in Figure~\ref{maxH2-B}. Note that in
runs B4, B7, and B8, $f_{\rm cc}$ remains zero until the end of the
run.}
\label{fcc-B}
\end{figure}

Given the role played by the high fractional ionization in boosting
$\mHt$ formation in this set of runs, we might expect the
uncertainties in the associative detachment and mutual neutralization
rate coefficients to have far more effect here than they did in runs
A1-A9. Indeed, this is what we find. In
Figures~\ref{maxH2-B}-\ref{minT-B}, we show how the central $\mHt$
abundance, central gas density, and central gas temperature vary over
the course of runs B1-B9, while in Figure~\ref{fcc-B} we plot the
evolution with time of $f_{\rm cc}$, the fraction of the gas which is
cool and condensed, defined here as the fraction of the total gas
which has a temperature $T < 500 \: {\rm K}$ and a density $n > 100 \:
{\rm cm^{-3}}$.

Figure~\ref{maxH2-B} demonstrates that the differences in the
associative detachment and mutual neutralization rate coefficients
used in these runs lead to large differences in the resulting $\mHt$
abundances. For instance, the difference between the value of $x_{\rm
\mHt, c}$ in runs B3 and B7 at $t = 150 \: {\rm Myr}$ is greater
than an order of magnitude; and while there is some indication that
the value of $x_{\rm \mHt, c}$ in the different runs begins to
converge at late times, the differences between the runs remain
significant at the end of the simulations.  It should also be noted
that the sense of the differences is precisely what we would expect,
based on our previous discussion of the chemistry: $\mHt$ formation is
more efficient in runs with a large value of the associative
detachment rate coefficient than in runs with a smaller value, but is
less effective when the mutual neutralization rate coefficient
is large than when it is small.

Figures~\ref{maxn-B} and \ref{minT-B} follow the evolution of
the density and temperature at the center of the protogalaxy and
demonstrate the effect that these differences in $\mHt$ abundance have
on the thermal and dynamical evolution of the gas. For the first $70
\: {\rm Myr}$ there is very little difference between the runs. In
each case, the gas initially cools to $T \sim 4000 \: {\rm K}$ through
a combination of Lyman-$\alpha$ cooling (at $T > 8000 \: {\rm K}$) and
Compton cooling, before gradually being re-heated by adiabatic
compression in the gravitational potential well of the dark matter. At
later times, the gas begins to cool more strongly through $\mHt$
ro-vibrational emission, and the ensuing loss of pressure support
leads to the gas rapidly collapsing to high densities. However, the
time of onset of this phase of cooling and collapse is sensitive to
the $\mHt$ abundance of the gas, and so occurs significantly later in
runs with small $\mHt$ abundances than in runs with high $\mHt$
abundances.  The physical reason for this dependence is clear: since
$\mHt$ is the dominant coolant in the gas at $t = 70 \: {\rm Myr}$,
the cooling time of the gas scales roughly as $t_{\rm cool} \propto
x_{\mHt}^{-1}$, and so protogalaxies in which the gas forms less
$\mHt$ naturally take longer to cool.

One important consequence of this is that the runs display a wide
variation in the amount of cold dense gas available for star
formation. This is illustrated by our plot of $f_{\rm cc}$ in
Figure~\ref{fcc-B}, which shows that the fraction of cold, dense gas
varies from 0\% (in runs B4, B8 \& B9) to almost 25\% (in run
B3). Given a sufficiently long time to evolve, it is likely that a
significant amount of cold gas will eventually accumulate in the runs
with less efficient $\mHt$ formation. However, we must recall that
unlike our idealized model protogalaxies, real protogalaxies are not
isolated systems, and do not have an unlimited time in which to evolve
before being disrupted by mergers or other external events (such as
nearby galactic outflows; see \citealt{tsd02}). Therefore, our
determination of whether or not a given halo can cool quickly enough
to form stars may depend on our choice of associative detachment and
mutual neutralization rate coefficients.

\subsection{Cooling and collapse with a UV background}
\label{hot_uv}
So far, we have assumed that the effects of any external UV background 
are negligible. In practice, this is unlikely to be the case -- by 
$z = 20$ a considerable UV background may already have developed 
\citep{har00,gb03}. Indeed, if cosmological
reionization is to occur somewhere in the redshift range $z_{\rm
reion} = 17 \pm 5$, as is suggested by the {\sc wmap} polarization
results \citep{wmap03}, there must already be a fairly strong
background in place. To explore how the presence of a UV background
may influence our conclusions, we have run several sets of simulations
using the same initial conditions as runs B1-B9, but in which the
strength of the UV background was varied as listed in
Table~\ref{runtab}.
 
\begin{figure}
\centering
\epsfig{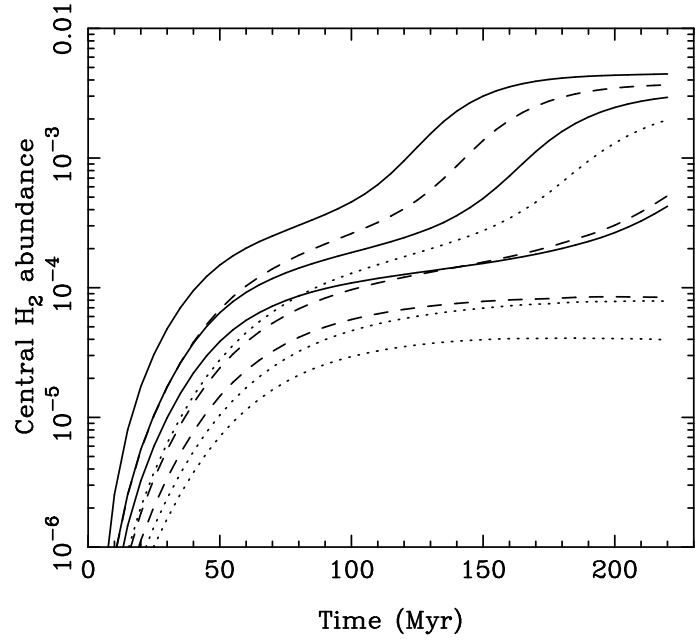}
\caption{The evolution with time of the central $\mHt$ abundance in runs
C1-C9.  The ordering of the lines is the same as in
Figure~\ref{maxH2-B}.}
\label{maxH2-001}
\end{figure}

\begin{figure}
\centering
\epsfig{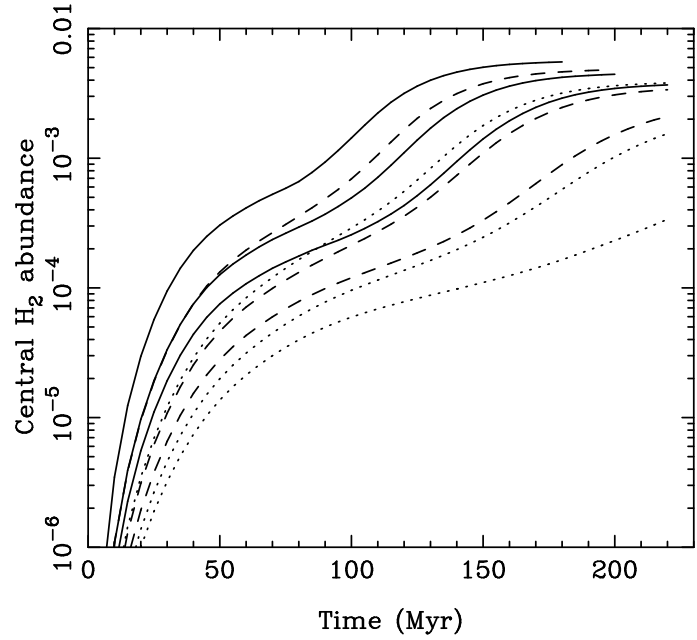}
\caption{Same as Figure~\ref{maxH2-001}, but for runs D1-D9.}
\label{maxH2-0003}
\end{figure}

\begin{figure}[Ht]
\centering
\epsfig{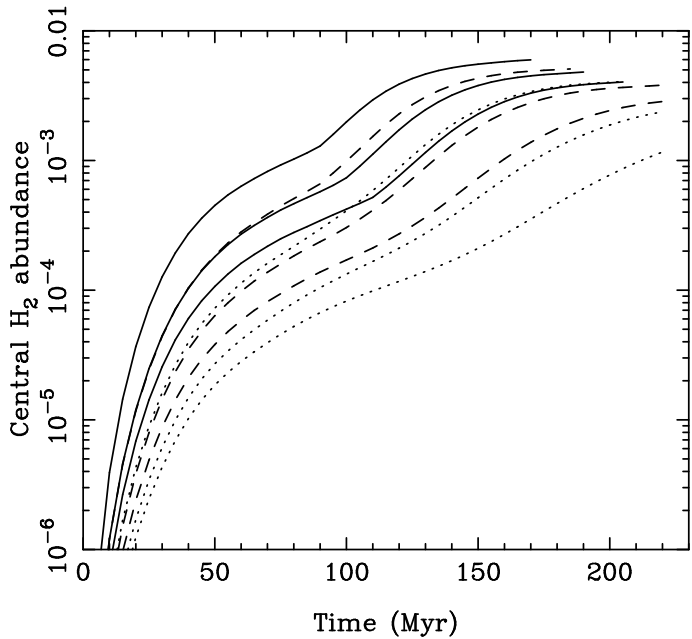}
\caption{Same as Figure~\ref{maxH2-001}, but for runs E1-E9.}
\label{maxH2-0001}
\end{figure}

In Figures~\ref{maxH2-001}--\ref{maxH2-0001}, we show how the central
$\mHt$ abundance varies with time in runs C1-C9, D1-D9 and E1-E9,
which correspond to UV field strengths specified by $J_{21} =
10^{-2}$, $3 \times 10^{-3}$ and $10^{-3}$ respectively. From
Figure~\ref{maxH2-001}, it is apparent that the presence of a strong
UV background significantly enlarges the difference in outcomes
between runs with different associative detachment and mutual
neutralization rate coefficients. For instance, the central $\mHt$
abundance at the end of run C3 differs from that at the end of run C7
by almost a factor of 100, whereas runs B3 and B7 differ in this
respect by no more than a factor of 4. However, from
Figures~\ref{maxH2-0003} \& \ref{maxH2-0001} we see that as the field
weakens, the differences become less pronounced; indeed, the results
of runs E1-E9 are close to those of runs B1-B9.

The reason for the large spread in outcomes that we see in runs C1-C9
becomes clearer once we consider that photodissociation by the imposed
UV background will set an upper limit on the $\mHt$ abundance in
optically thin gas, given by the equilibrium value at which formation
balances photodissociation:
\begin{equation}
x_{\rm \mHt, eq} \simeq \frac{R_{\mHt}}{R_{\rm pd}} x_{\mH}.
\end{equation}
Here $R_{\rm pd}$ is the photodissociation rate per $\mHt$ molecule,
and $R_{\mHt}$ is the rate of $\mHt$ formation per H atom. To compute 
the latter, we note that
the dominant contribution comes from $\mHt$ formation via the  $\Hm$ pathway, 
and so we can approximate $R_{\mHt}$ by the product of the $\Hm$ formation 
rate per $\mH$ atom and the fraction of $\Hm$ that successfully forms $\mHt$ 
(rather than being destroyed by mutual neutralization). Therefore
\begin{equation}
 R_{\mHt} = k_{\Hm} n_{\me} \times 
\frac{k_{\rm ad} n_{\mH}}{k_{\rm ad} n_{\mH} + k_{\rm mn} n_{\Hp}},
\end{equation}
where $k_{\Hm}$ is the rate coefficient for the formation of $\Hm$ by
radiative association, and where $k_{\rm ad}$ and $k_{\rm mn}$ are the
associative detachment and mutual neutralization rate coefficients
respectively. In gas with a high fractional ionization, 
$k_{\rm mn} n_{\Hp} \gg k_{\rm ad} n_{\mH}$, and so this 
expression reduces to
\begin{equation}
R_{\mHt} \simeq k_{\Hm} \frac{k_{\rm ad}}{k_{\rm mn}} 
\frac{n_{\me}}{n_{\Hp}} n_{\mH},
\end{equation}
which can be further simplified to 
\begin{equation}
R_{\mHt} \simeq k_{\Hm} \frac{k_{\rm ad}}{k_{\rm mn}} n_{\mH},
\end{equation}
provided that $n_{\me} \simeq n_{\Hp}$. After $100 \: {\rm Myr}$ of
evolution, the gas at the center of our simulated protogalaxies in
runs C1-C9 has a temperature of approximately $T = 7100 \: {\rm K}$
and an atomic hydrogen number density of approximately $n_{\mH} = 2.6
\times 10^{-2} \: {\rm cm^{-3}}$; the variation in these values
between the different runs is about 5--10\%. Using these values for
$n_{\mH}$ and $T$, we find that $k_{\Hm} = 3 \times 10^{-15} \: {\rm
cm^{3}} \: {\rm s^{-1}}$ \citep{wish79}, and hence that $R_{\mHt} = 3
\times 10^{-15} (k_{\rm ad} / k_{\rm mn}) n_{\mH} \:
\rm{s}^{-1}$. Now, since $R_{\rm pd} = 1.3 \times 10^{-14} \: {\rm
s^{-1}}$, given our assumed UV field strength of $J_{21} = 10^{-2}$
\citep{db96}, it is easy to show that
\begin{equation}
x_{\mHt, {\rm eq}} \simeq 6.0 \times 10^{-3} 
\left(\frac{k_{\rm ad}}{k_{\rm mn}}
\right),
\end{equation}
where the size of the term in brackets varies from $5 \times 10^{-3}$
to $6 \times 10^{-1}$ depending on the values chosen for the rate
coefficients. The equilibrium $\mHt$ abundance in optically thin gas
therefore varies from $3.1 \times 10^{-5}$ to $3.6 \times 10^{-3}$, 
depending on our choice of chemical rate coefficients. If we take the
effects of self-shielding into account, we will obtain slightly larger
values, but if anything the spread in outcomes will be even greater, 
since gas with a large $\mHt$ abundance can self-shield far more effectively 
than gas with a small $\mHt$ abundance.  

\begin{figure}[Ht]
\centering
\epsfig{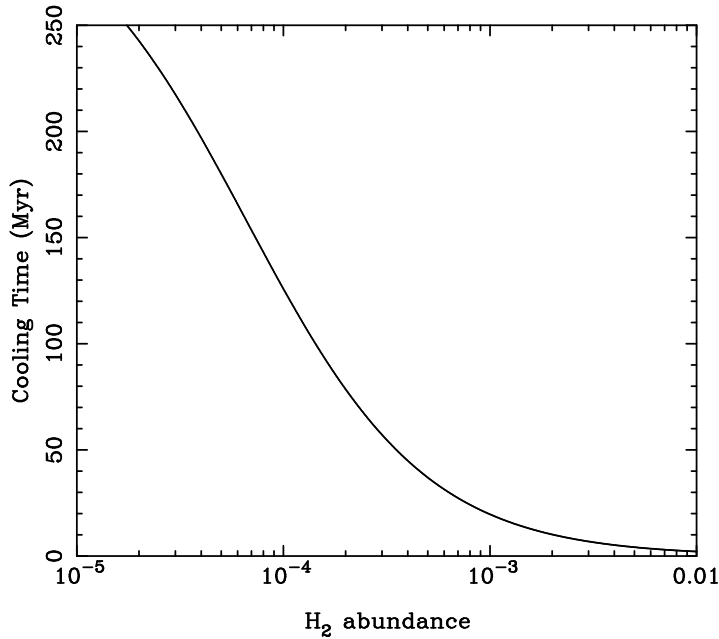}
\caption{The cooling time of gas as a function of its $\mHt$
abundance, computed assuming a temperature $T = 7100 \: {\rm K}$ 
and density $n_{\mH} = 2.6 \times 10^{-2} \: {\rm cm^{-3}}$. 
These values are representative of the central temperatures and
densities found within the halos simulated in runs C1-C9.  Values
in the individual runs differ slightly from these values, but never
by more than 5-10\%.}
\label{tcool}
\end{figure}

\begin{figure}
\centering
\epsfig{figure=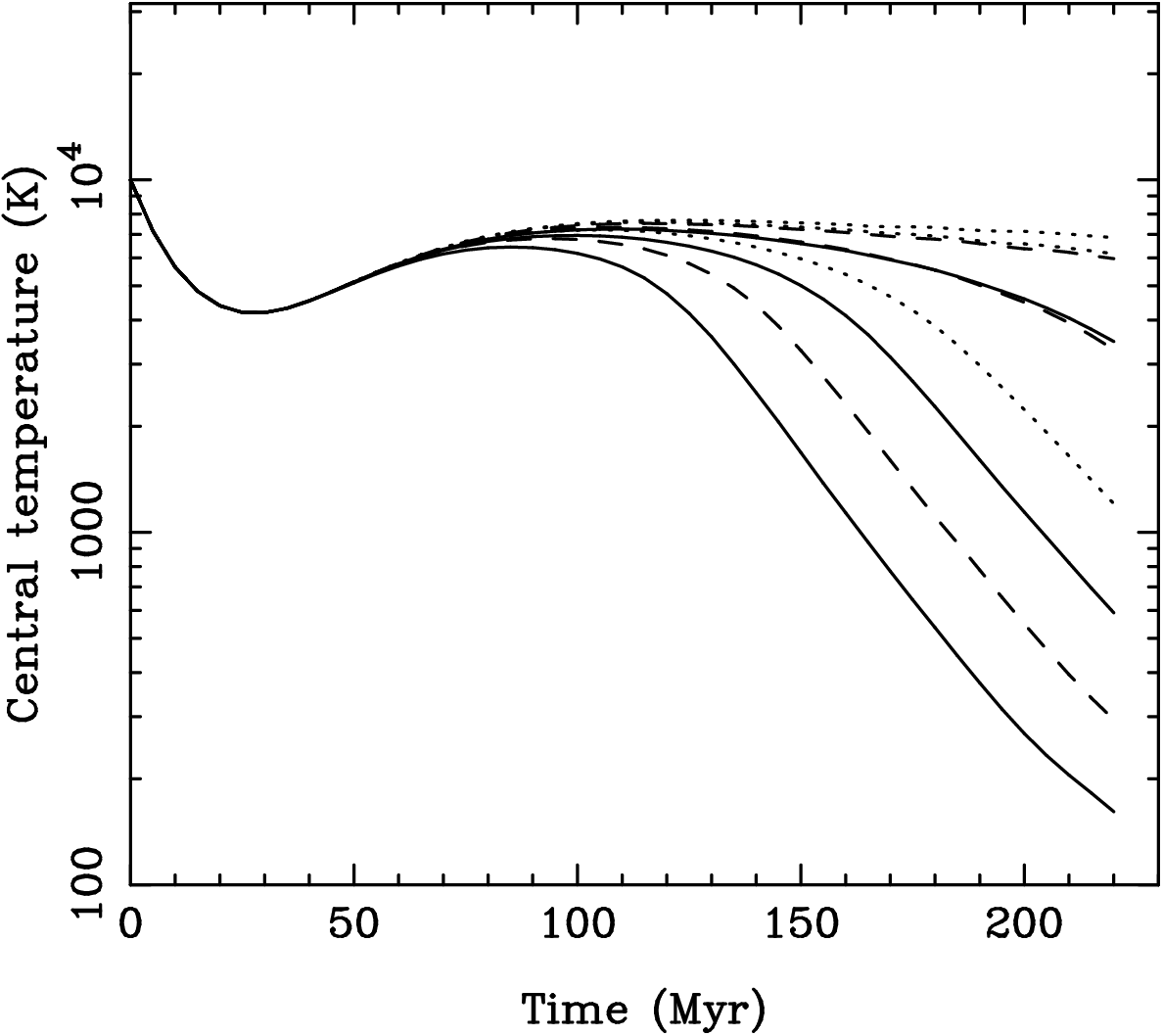,width=20pc,angle=270,clip=}
\caption{Same as Figure~\ref{minT-B}, but for runs C1-C9.}
\label{minT-001}
\end{figure}

\begin{figure}
\centering
\epsfig{figure=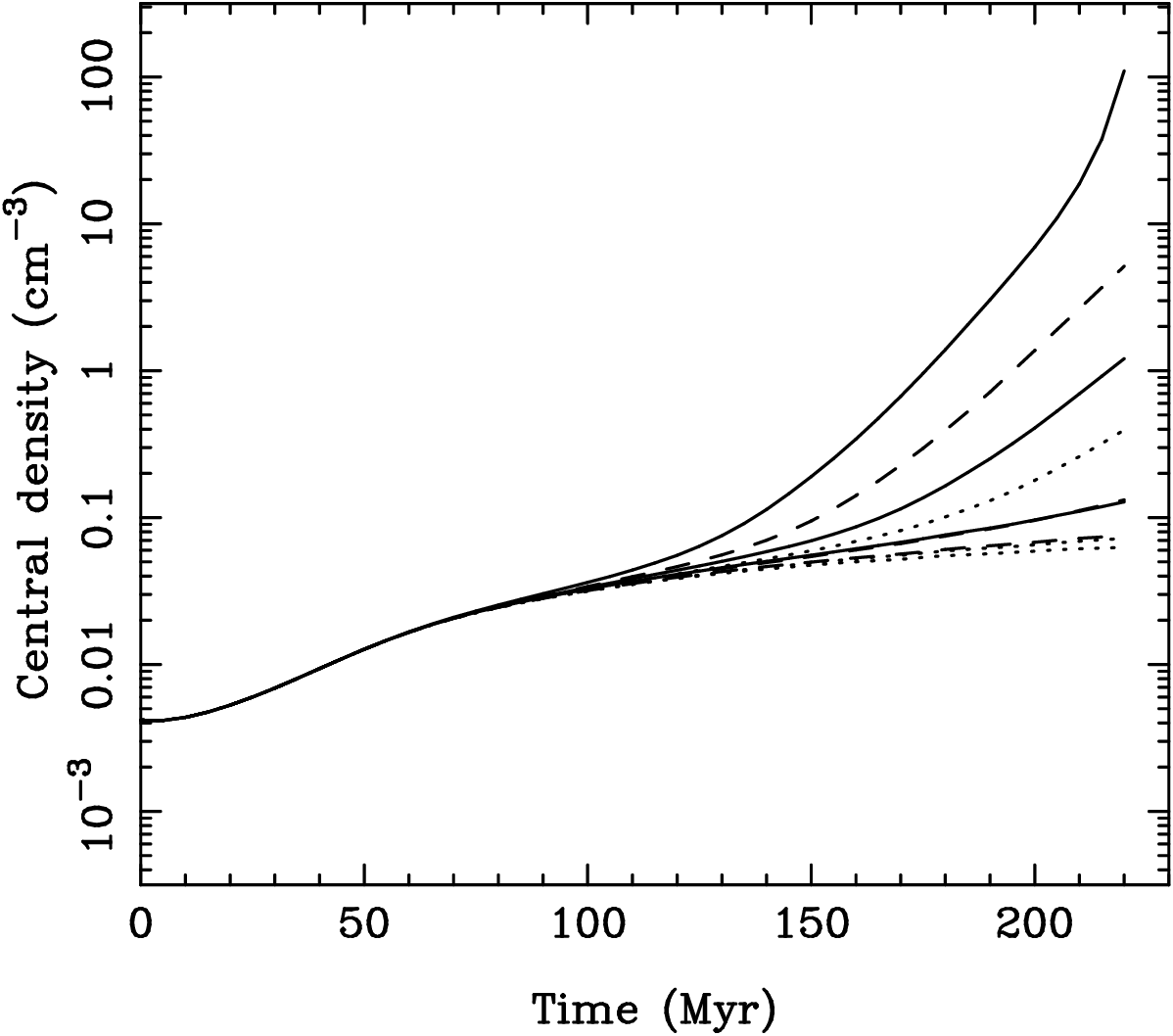,width=20pc,angle=270,clip=}
\caption{Same as Figure~\ref{maxn-B}, but for runs C1-C9.}
\label{maxn-001}
\end{figure}

If we now compute the cooling timescale for this gas, which is given
by
\begin{equation}
t_{\rm cool} \simeq \frac{\frac{3}{2} n_{\rm tot} k T}{\Lambda},
\end{equation}
where $n_{\rm tot}$ is the total number density of particles
(including electrons), and where $\Lambda$ is the cooling rate,
computed using the SPH code in units of ${\rm erg \: cm^{-3} \:
s^{-1}}$ for a range of different $\mHt$ abundances (see
Figure~\ref{tcool}), then we can determine how much $\mHt$ is required
in order to cool the gas before the end of the simulation.  From
Figure~\ref{tcool}, we see that an $\mHt$ abundance $x_{\mHt} \gtrsim
10^{-4}$ is required (as here we are at 100 Myr and the simulations
run for 220 Myr). Therefore, for those of our runs in which $x_{\rm
\mHt, eq}$ is larger than this value, we would expect the gas to cool
efficiently, while in runs in which $x_{\rm \mHt, eq}$ is not
sufficiently large, we would expect the gas to cool slowly, if at
all. If we examine the evolution of the temperature of the gas at the
center of the protogalaxies simulated in these runs, which we plot in
Figure~\ref{minT-001}, we find that our results are consistent with
these expectations: in runs C6, C8 \& C9, where $x_{\rm \mHt, eq} <
10^{-4}$, very little cooling occurs, while in the remainder of the
runs, which have $x_{\rm \mHt, eq} > 10^{-4}$, far more cooling
occurs. It is also clear from the plot that the final temperature
reached in the runs that cool depends strongly on the value of $x_{\rm
\mHt, eq}$. The differences between the thermal evolution of the gas
in the various runs are also reflected in the dynamical evolution, as
can be seen from Figure~\ref{maxn-001}.  This example therefore gives
a particularly striking demonstration of the need for better rate
coefficient data, since without it we cannot determine which of the
various outcomes illustrated in Figures~\ref{maxH2-001},
\ref{minT-001} \& \ref{maxn-001} is actually the correct one.

It is also interesting to compare the length of the cooling time in these 
runs with the free-fall time, given by
\begin{equation}
t_{\rm ff} = \sqrt{\frac{3\pi}{32G\rho_{\rm tot}}},
\end{equation}
where $\rho_{\rm tot}$ is the total matter density (i.e., the sum of
the gas density and the dark matter density). At this point of 100 Myr
in the simulation, the main contribution to $\rho_{\rm tot}$ comes
from the dark matter.  Evaluating $t_{\rm ff}$ at the center of the
dark matter halo, we find that is has a value of approximately $20 \:
{\rm Myr}$. From Figure~\ref{tcool}, it is clear that for the gas to
have $t_{\rm cool} < t_{\rm ff}$, it must have an $\mHt$ abundance
$x_{\mHt} \gtrsim 10^{-3}$. Since this value is an order of magnitude
larger than the abundance required to allow cooling by the end of the
simulation, i.e., within what is essentially a Hubble time $t_{\rm
H}$, it is possible for there to be protogalaxies in which $t_{\rm ff}
< t_{\rm cool} < t_{\rm H}$. In these protogalaxies, the timescale on
which the gas collapses will be determined by $t_{\rm cool}$, rather
than $t_{\rm ff}$, and so will be directly sensitive to the amount of
$\mHt$ within the gas. Therefore, predictions that we make concerning
the dynamical evolution of such protogalaxies will be highly sensitive
to our choices for the mutual neutralization and associative
detachment rate coefficients, as can already been seen from
Figure~\ref{maxn-001}.

\begin{figure}
\centering
\epsfig{figure=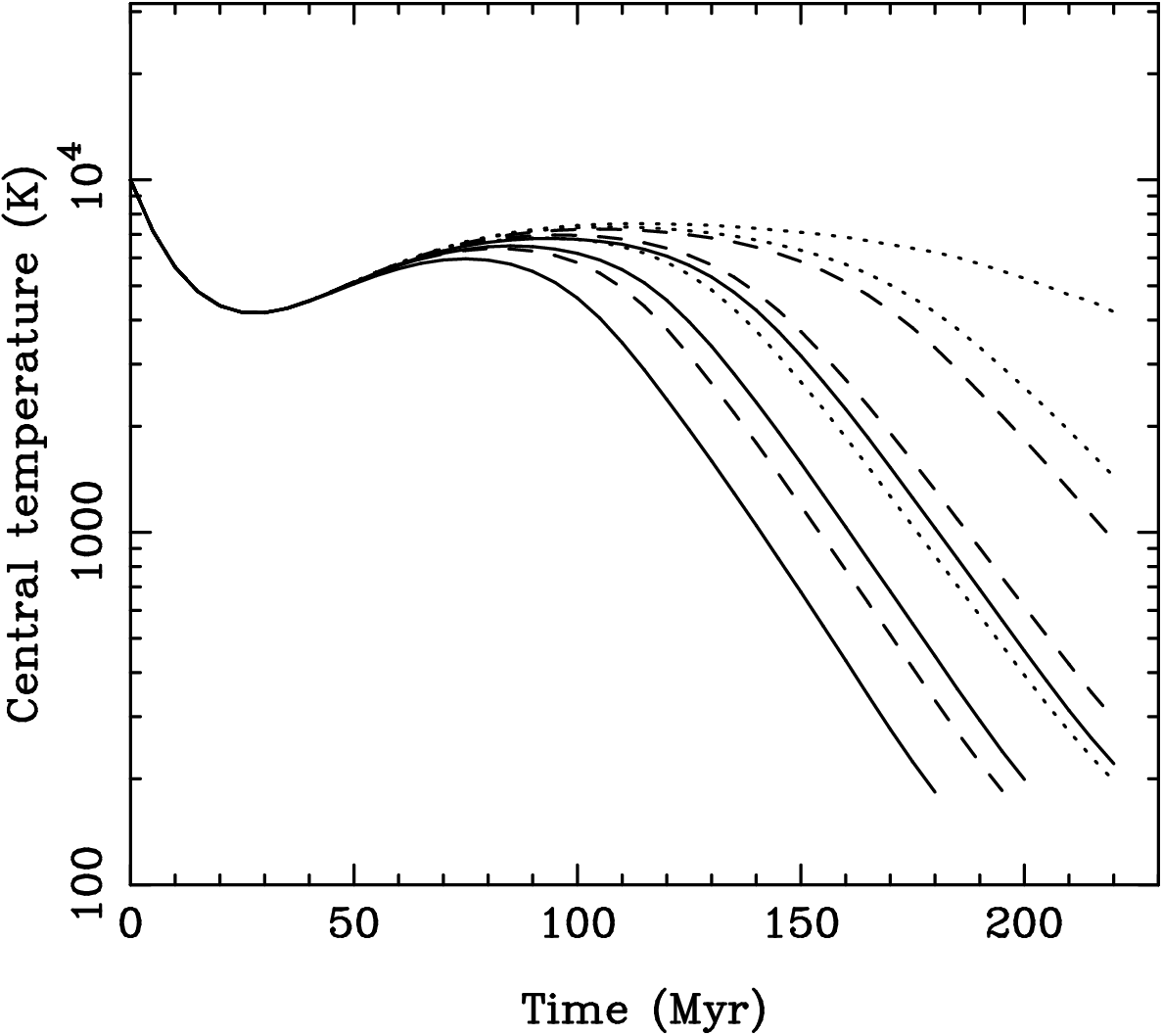,width=20pc,angle=270,clip=}
\caption{Same as Figure~\ref{minT-B}, but for runs D1-D9.}
\label{minT-0003}
\end{figure}

\begin{figure}
\centering
\epsfig{figure=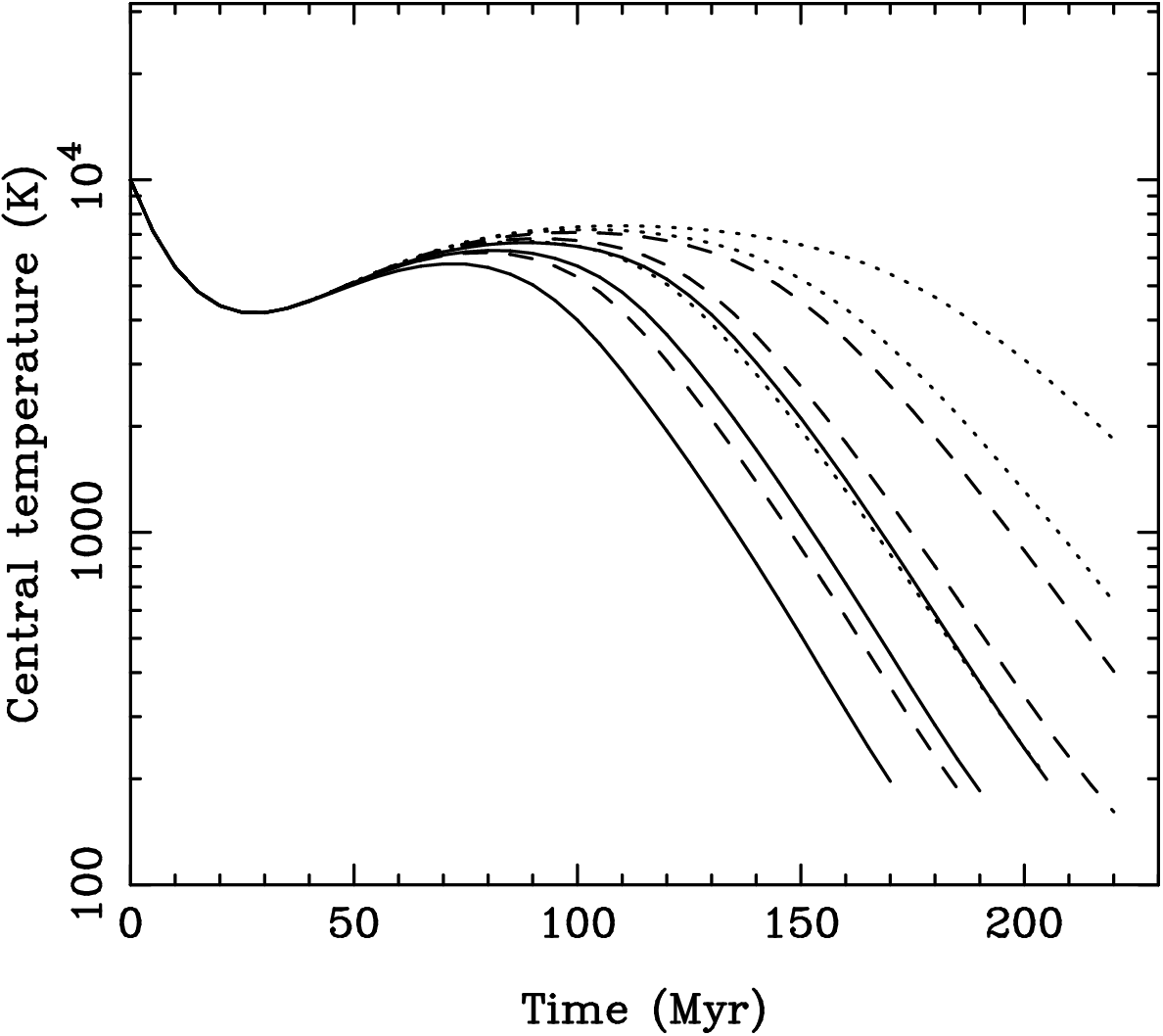,width=20pc,angle=270,clip=}
\caption{Same as Figure~\ref{minT-B}, but for runs E1-E9.}
\label{minT-0001}
\end{figure}

If the strength of the UV background is reduced, as in runs D1-D9 or
E1-E9, then the value of $x_{\rm \mHt, eq}$ increases, and so the
uncertainty in the rate coefficients has less influence, as it is
easier for the gas to form enough $\mHt$ to provide efficient
cooling. Nevertheless, the differences between the runs remain
significant, as can be seen from the plots of $\mHt$ abundance versus
time in Figures~\ref{maxH2-0003} and \ref{maxH2-0001}, or the plots of
temperature versus time shown in Figures~\ref{minT-0003} and
\ref{minT-0001}.

One final comparison that we can make between the runs is to examine
how sensitive $f_{\rm cc}$ is to the values of the associative
detachment and mutual neutralization rate coefficients, and to the
strength of the UV background.  In Table~\ref{fcc_tab} we list the
value of $f_{\rm cc}$ at the end of the simulation for all of the runs
in sets B, C, D and E. The trend of increasing $f_{\rm cc}$ with
decreasing field strength is clear, and is easily understood given our
discussion above. However, the table also demonstrates the degree of
variation in $f_{\rm cc}$ that is introduced by the uncertainties in
the rate coefficients. In particular, we see that the lack of
reliability in the atomic data introduce an uncertainty of almost an
order of magnitude into the value of the UV field strength that is
required to prevent gas cooling and collapse.

\section{Discussion}
\label{discuss}

The results that we have presented in this paper serve as a
demonstration of the potential impact that the existing uncertainties
in the values of the associative detachment and mutual neutralization
rate coefficients will have on our ability to make accurate
predictions regarding protogalactic cooling and star formation.

For protogalaxies forming from cold gas with a low initial fractional
ionization, the impact is small -- some uncertainty is introduced into
the predicted $\mHt$ abundance, but not enough to significantly alter
the dynamical evolution of the gas. If we were to continue our simulations 
for a longer period, then we would expect $\mHt$ formation in the 
cooling, dense gas at the centre of the protogalaxy to proceed at a very
similar rate in all of our runs.
 
For protogalaxies forming from hot, highly ionized gas, however, the
impact is substantial. The uncertainties in the chemical rate
coefficients are responsible for producing a large uncertainty in the
predicted $\mHt$ abundance, which in turn introduces a large
uncertainty into the cooling rate of the gas. Since the timescale
on which the gas collapses is set by the larger of $t_{\rm cool}$ and
$t_{\rm ff}$, and since, in most protogalaxies, $t_{\rm cool}$ is 
initially longer than $t_{\rm ff}$, any uncertainty in the cooling 
rate has a direct impact on the dynamical evolution of the gas. Indeed, 
in the examples studied in this paper,
our choice of rate coefficients has a great influence on the
conclusion that we draw regarding whether or not the gas can cool and
collapse within the lifetime of the protogalaxy (which will be of the
order of a Hubble time). The presence of an ultraviolet background
only serves to exacerbate this problem, as it increases the
sensitivity of the predicted $\mHt$ abundance to our choice of
chemical rate coefficients.

Ultimately, the only way to remove these uncertainties will be to
obtain more accurate rate coefficients for the relevant associative
detachment and mutual neutralization processes.  Most important for
this will be new laboratory measurements for each reaction at
cosmologically relevant collision energies supplemented by further
theoretical calculations.

\acknowledgements

The authors would like to thank R. Klessen, M.-M. Mac Low and 
P.C. Stancil for helpful comments on an earlier draft of this paper.
S.C.O.G. is supported by NSF grant AST-0307793 and NASA Education grant
NAG5-13028. D.W.S. is supported in part by a NASA Space Astrophysics
Research and Analysis grant NAG5-5420, a NASA Astrophysics Theory
Program grant NNG04GL39G, and an NSF Galactic Astronomy Program grant
AST-03-7203.  A.K.J. acknowledges support from the Emmy Noether
Program of the Deutsche Forschungsgemeinschaft (grant no. KL1358/1).

\clearpage

\begin{deluxetable}{lll}
\tablewidth{0pt}
\tablecaption{A list of all the reactions included in our model 
of primordial gas chemistry. \label{chemtab}}
\tablehead{\multicolumn{1}{l}{Reaction Number} & 
\multicolumn{1}{l}{Reaction} & 
\multicolumn{1}{l}{Reference}}
\startdata
1 & $\mH  + \me  \rightarrow \Hm + \gamma$ & \citet{wish79} \\
& \\
2 & $\Hm  + \mH  \rightarrow \mHt + \me$ & {\rm See text} \\
& \\
3 & $\mH  + \Hp  \rightarrow \mHtp + \gamma$ & \citet{rp76} \\
& \\
4 & $\mH + \mHtp \rightarrow \mHt + \Hp$ & \citet{kah79} \\
& \\
5 & $\Hm  + \Hp  \rightarrow \mH + \mH$ & {\rm See text} \\
& \\
6 & $\Hm + \gamma \rightarrow \mH + \me$ & \citet{wish79} \\
& \\
7 & $\mHtp + \me \rightarrow \mH + \mH$ & \citet{sdgr94} \\
& \\
8 & $\mHt + \Hp  \rightarrow \mHtp + \mH$ & \citet{skhs04} \\
& \\
9 & $\mHt + \me  \rightarrow  \mH + \mH +  \me$ & \citet{st99} \\
& \\
10 & $\mHt + \mH  \rightarrow  \mH + \mH + \mH$  & \citet{ms86} \\
& \\
11 & $\mHt + \gamma \rightarrow \mH + \mH$ & \citet{db96} \\
& \\
12 & $\mH  + \me  \rightarrow \Hp + \me + \me$ & \citet{janev87} \\
& \\
13 & $\Hp  + \me  \rightarrow \mH +  \gamma$ & \citet{fer92} \\
& \\
14 & $\Hm  + \me  \rightarrow \mH + \me + \me$ & \citet{janev87} \\
& \\
15 & $\Hm  + \mH  \rightarrow  \mH + \mH +  \me$ & \citet{janev87} \\
& \\
16 & $\Hm + \Hp   \rightarrow \mHtp + \me$ & \citet{pbcmv78} \\
& \\
17 & $\mHtp + \gamma \rightarrow \mH + \Hp$ & \citet{dunn68} \\
& \\
\enddata
\tablerefs{References are to the primary source of data for each
reaction. Photochemical reactions assume an incident spectrum
corresponding to a modified, diluted black-body, as described in the
text.}
\end{deluxetable}

\begin{deluxetable}{ccccc}
\tablewidth{0pt}
\tablecaption{Initial conditions for the simulations.
\label{runtab}}
\tablehead{\colhead{Run} & \colhead{$T_{\rm g}$ (K) } &
\colhead{$x_{\me}$} & \colhead{$x_{\mHt}$} &
\colhead{$J_{21}$}}
\startdata
A & 12 & $2.2 \times 10^{-4}$  & $2.4 \times 10^{-6}$  & 0.0 \\
B & $10^{4}$ & 1.0 & 0.0 & 0.0 \\
C & $10^{4}$ & 1.0 & 0.0 & $10^{-2}$ \\
D & $10^{4}$ & 1.0 & 0.0 & $3 \times 10^{-3}$ \\
E & $10^{4}$ & 1.0 & 0.0 & $10^{-3}$ \\
\enddata
\end{deluxetable}

\begin{deluxetable}{ccc}
\tablewidth{0pt}
\tablecaption{The values of the mutual neutralization and associative
detachment rate coefficients used in our runs. We list here all nine 
possible combinations, together with the number used elsewhere 
in the paper to refer to each combination. 
\label{settab}}
\tablehead{\colhead{Set} & \multicolumn{2}{c}{Rate 
coefficient (cm$^3$~s$^{-1}$)}\\
\cline{2-3}
\colhead{} & \colhead{Mutual Neutralization} 
&\colhead{Associative Detachment}}
\startdata
1 & $7 \times 10^{-7} T^{-1/2}$ & $0.65 \times 10^{-9}$ \\
\\
2 & $7 \times 10^{-7} T^{-1/2}$ & $1.30 \times 10^{-9}$ \\
\\
3 & $7 \times 10^{-7} T^{-1/2}$ & $5.00 \times 10^{-9}$ \\
\\
4 & $2.4 \times 10^{-6} (1 + 5 \times 10^{-5} T) T^{-1/2}$
  & $0.65 \times 10^{-9}$ \\ 
\\
5 & $2.4 \times 10^{-6} (1 + 5 \times 10^{-5} T) T^{-1/2}$ 
& $1.30 \times 10^{-9}$ \\
\\
6 & $2.4 \times 10^{-6} (1 + 5 \times 10^{-5} T) T^{-1/2}$ 
& $5.00 \times 10^{-9}$ \\
\\
7 & $\begin{array}{rcl}
     5.7 \times 10^{-6} T^{-1/2} & + & 6.3 \times 10^{-8} \\ 
   - 9.2 \times 10^{-11} T^{1/2} & + & 4.4 \times 10^{-13} T
     \end{array}$ & $0.65 \times 10^{-9}$ \\
\\
8 & $\begin{array}{rcl}
     5.7 \times 10^{-6} T^{-1/2} & + & 6.3 \times 10^{-8} \\ 
   - 9.2 \times 10^{-11} T^{1/2} & + & 4.4 \times 10^{-13} T
     \end{array}$ & $1.30 \times 10^{-9}$ \\
\\
9 & $\begin{array}{rcl}
     5.7 \times 10^{-6} T^{-1/2} & + & 6.3 \times 10^{-8} \\ 
   - 9.2 \times 10^{-11} T^{1/2} & + & 4.4 \times 10^{-13} T
     \end{array}$ & $5.00 \times 10^{-9}$ \\
\enddata
\tablerefs{The mutual neutralization rate coefficients are taken from
\citet{dl87}, \citet{cdg99} and \citet{pams71} respectively. For the
associative detachment rate coefficients, the central value is taken
from \citet{sff67}; the other values represent lower and upper bounds
on plausible values.}
\end{deluxetable}

\begin{deluxetable}{ccccc}
\tablewidth{0pt}
\tablecaption{Physical state of the densest gas at the end of runs A1-A9.
\label{run_a_results}}
\tablehead{\colhead{Run} & \colhead{$T_{\rm c} /  10^{3} \: {\rm K}$} &
 \colhead{$n_{\rm c}  / 10^{-2} \: {\rm cm^{-3}}$} &
 \colhead{$x_{\rm \me, c} / 10^{-4}$} & \colhead{$x_{\rm \mHt, c} / 10^{-5}$}}
\startdata
A1 & $6.94 $ & $6.47 $ & $1.98 $ & $1.83 $ \\
A2 & $6.66 $ & $6.75 $ & $1.98 $ & $3.08 $ \\
A3 & $6.30 $ & $7.30 $ & $1.98 $ & $5.46 $ \\
A4 & $7.72 $ & $6.54 $ & $1.98 $ & $0.66$ \\
A5 & $7.44 $ & $6.71 $ & $1.98 $ & $1.24$ \\
A6 & $6.35 $ & $7.32 $ & $1.98 $ & $5.06$ \\
A7 & $6.92 $ & $6.48 $ & $1.96 $ & $1.85 $ \\
A8 & $6.66 $ & $6.79 $ & $1.96 $ & $3.00 $ \\
A9 & $6.30 $ & $7.25 $ & $1.97 $ & $5.17 $ \\
\enddata
\end{deluxetable}

\begin{deluxetable}{ccccc}
\tablewidth{0pt}
\tablecaption{The value of the cold, collapsed gas fraction, $f_{\rm cc}$, at
the end of the simualtions. \label{fcc_tab}}
\tablehead{\colhead{Run} & \colhead{$J_{21} = 10^{-2}$} &
 \colhead{$J_{21} = 3 \times 10^{-3}$} &
 \colhead{$J_{21} = 10^{-3}$} & \colhead{$J_{21} = 0.0$}}
\startdata
1 & $0.00 $ & $0.00 $ & $0.08 $ & $0.12 $ \\
2 & $0.00 $ & $0.08 $ & $0.14 $ & $0.17 $ \\
3 & $2 \times 10^{-4} $ & $0.15 $ & $0.19 $ & $0.23 $ \\
4 & $0.00 $ & $0.00$  & $0.00 $ & $0.00$ \\
5 & $0.00 $ & $0.00 $ & $0.005 $ & $0.08$ \\
6 & $0.00 $ & $0.10 $ & $0.15 $ & $0.18$ \\
7 & $0.00 $ & $0.00 $ & $0.00 $ & $0.00 $ \\
8 & $0.00 $ & $0.00 $ & $0.00 $ & $0.00 $ \\
9 & $0.00 $ & $0.00 $ & $0.07 $ & $0.11 $ \\
\enddata
\end{deluxetable}

\end{document}